\documentclass{article}

\usepackage{changepage}

\usepackage[main, final, nonatbib]{iaseai26}
\usepackage[authoryear, round]{natbib}

\usepackage{pdflscape}      
\usepackage[utf8]{inputenc} 
\usepackage[T1]{fontenc}    
\usepackage{hyperref}       
\usepackage{url}            

\usepackage{booktabs}       
\usepackage{amsfonts}       
\usepackage{nicefrac}       
\usepackage{microtype}      
\usepackage{xcolor}         
\usepackage[most]{tcolorbox} 
\usepackage{graphicx}
\usepackage{float}
\usepackage{longtable} 
\usepackage{caption}
\usepackage{ltablex} 
\usepackage{array}
\usepackage{ragged2e}
\usepackage{enumitem}

\usepackage[capitalise]{cleveref}

\usepackage[toc,page]{appendix}
\crefname{appendix}{Appendix}{Appendices} 
\Crefname{appendix}{Appendix}{Appendices}

\keepXColumns

\title{The Role of Risk Modeling in Advanced AI Risk Management}

 \author{%
Chloé Touzet$^{1,}$\thanks{Corresponding author \texttt{chloe@safer-ai.org}} \quad 
\textbf{Henry Papadatos$^1$} \quad \textbf{Malcolm Murray$^1$} \quad \textbf{Otter Quarks$^1$} \quad \textbf{Steve Barrett$^1$}\\
\textbf{Alejandro Tlaie Boria$^1$} \quad \textbf{Elija Perrier$^2$} \quad \textbf{Matthew Smith$^1$}\quad \textbf{Siméon Campos$^1$}\\
\\
$^1$SaferAI \quad $^2$University of Technology Sydney \\
}

\begin{document}

\maketitle

\begin{abstract}
  Rapidly advancing artificial intelligence (AI) systems introduce novel, uncertain, and potentially catastrophic risks. Managing these risks requires a mature risk-management infrastructure whose cornerstone is rigorous risk modeling. We conceptualize AI risk modeling as the tight integration of (i) scenario building---causal mapping from hazards to harms---and (ii) risk estimation---quantifying the likelihood and severity of each pathway. We review classical techniques such as Fault and Event Tree Analyses, FMEA/FMECA, STPA and Bayesian networks, and show how they can be adapted to advanced AI. A survey of emerging academic and industry efforts reveals fragmentation: capability benchmarks, safety cases, and partial quantitative studies are valuable but insufficient when divorced from comprehensive causal scenarios. Comparing the nuclear, aviation, cybersecurity, financial, and submarine domains, we observe that every sector combines deterministic guarantees for unacceptable events with probabilistic assessments of the broader risk landscape. We argue that advanced-AI governance should adopt a similar dual approach and that verifiable, provably-safe AI architectures are urgently needed to supply deterministic evidence where current models are the result of opaque end-to-end optimization procedures rather than specified by hand. In one potential governance-ready framework, developers conduct iterative risk modeling and regulators compare the results with predefined societal risk tolerance thresholds. The paper provides both a methodological blueprint and opens a discussion on the best way to embed sound risk modeling at the heart of advanced-AI risk management.
\end{abstract}

\begin{figure}[H]
  \centering
  \includegraphics[width=0.8\textwidth]{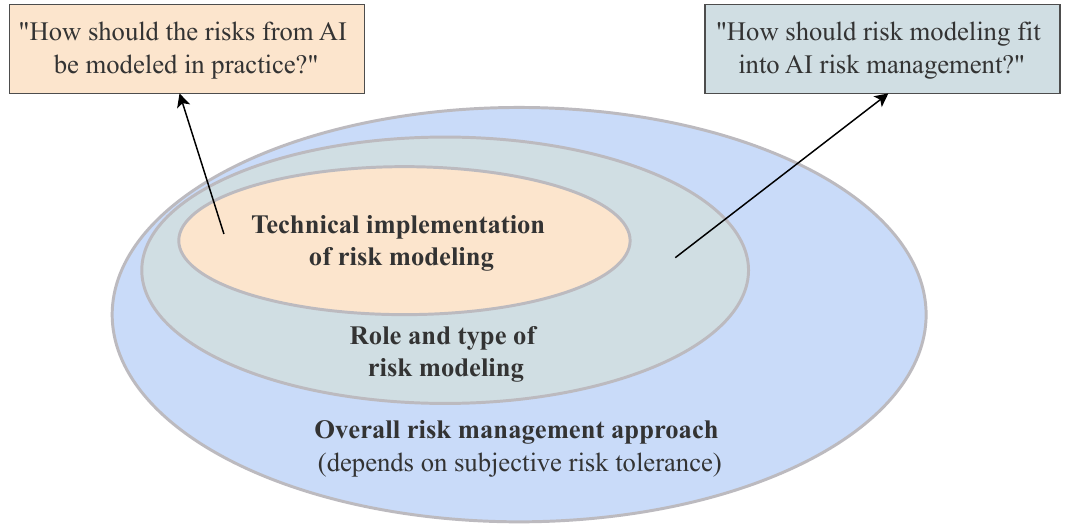}
  \label{Figure_circles}
\end{figure}

\newpage

\section*{Executive Summary / Highlights}
Rapidly advancing artificial intelligence (AI) systems introduce novel and potentially catastrophic risks, and they are being deployed amid deep epistemic uncertainty. \textbf{Safety-critical industries facing catastrophic hazards—such as nuclear power or aviation—have achieved dramatic safety gains by institutionalizing risk management, with rigorous risk modeling at its core.} Risk modeling is part of the explanation behind these industries’ improved safety.
 
In AI, practical risk modeling remains fragmented. \textbf{We define risk modeling as the tight coupling of (i) scenario building}, which maps causal pathways from hazard to harm, \textbf{and (ii) risk estimation}, which assigns likelihood and harm values to these scenarios, with explicit treatment of uncertainty and dependencies. \textbf{Both components are necessary}: estimation without scenarios cannot yield a comprehensive risk picture; scenarios without estimation cannot support real decision-making trade-offs.

This paper touches on three nested questions. The outer question is governance: \textit{\textbf{what risk-management approach should society adopt for advanced AI?}} This includes questions related to the roles of international bodies and national regulators, responsibility sharing with industry, transparency, and risk-tolerance setting. The middle question asks: \textit{\textbf{“how should risk modeling fit within that approach?"}} I.e., what blend of deterministic and probabilistic requirements, what concrete use of modeling outputs? The inner question is technical: \textit{\textbf{“how can risk modeling for advanced AI be done in practice?”}} This paper focuses on the inner two: it discusses how to adapt classical scenario building and risk estimation tools to advanced AI; it suggests one possible way to use risk modeling within risk management. It deliberately leaves final choices about institutional design and risk tolerance to policymakers, while making explicit the decisions they must settle. 

\textbf{On the technical question}, we (i) translate foundational risk-modeling concepts to AI contexts; (ii) adapt scenario-building tools (FTA/ETA, FMEA/FMECA, STPA, bow-tie) to AI scenarios; (iii) review quantitative techniques (structured expert elicitation, Monte Carlo, Bayesian Networks, copulas) and show how to connect them to advanced AI scenarios; and (iv) survey emerging AI-specific practices and gaps. Two principles recur: integration over isolation—\textbf{scenarios should be built to enable quantification, and quantification should respect scenario logic and dependencies}; and rigor over impressionism—use \textbf{structured elicitation with calibration and report uncertainty explicitly}. We distinguish safety cases (argumentative assurance) from comprehensive scenario modeling and argue that risk models should feed—rather than be replaced by—safety cases. Given heavy tails, sparse data, and rapid change, modeling must be dynamic and iterative, updating with evaluations, incidents, and red-team results.

On the question: \textbf{“how should risk modeling fit into advanced AI risk management”}, our \textbf{survey of five industries (nuclear, aviation, cybersecurity, finance, submarine operations) }yields two lessons. First, \textbf{mature sectors often mandate modeling aligned with international standards}; for AI, unresolved governance choices include who models, who audits, how results are shared, and which international bodies set norms. We illustrate one coherent option: regulators mandate scenario-based, dependency-aware modeling by developers; independent experts audit; regulators compare outputs to predefined risk tolerance thresholds in deployment certification. The second lesson is that \textbf{every sector blends probabilistic and deterministic elements. We argue that AI should do the same to meet safety-critical norms.} Yet AI’s intrinsic opacity hinders strong deterministic assurances,\textbf{motivating investment in verifiable AI safety (provable components, interpretable mechanisms) to enable hard guarantees for the highest-severity risks}.

This paper’s original contributions include: 
(1) an operationalization of AI risk modeling as coupled causal pathways and dependency-aware estimation; 
(2) an adaptation of classical tools (FTA/ETA, STPA; elicitation/Monte Carlo/Bayesian methods/BNs/copulas) to AI; 
(3) a clarification of safety-case limits and how explicit risk models should feed assurance; 
(4) a cross-industry map of modeling’s roles from conservative design margins to best-estimate profiles; 
(5) a suggested governance-ready framing that links model outputs to tolerability thresholds
(6) a case for research in verifiable AI safety to unlock deterministic guarantees despite black box systems.

Future work should prioritize three directions. 
\begin{itemize}
    \item First, we recommend \textbf{further technical developments to sharpen advanced AI risk methodology}, including \textbf{scalable, calibrated expert judgment}; \textbf{improved dependency and tail-risk methods}; \textbf{dynamic, iterative modeling with KRIs/KCIs}; and validated \textbf{mappings from lab capability evaluations to real-world risk}.
\end{itemize}
\begin{itemize}
    \item Second, \textbf{resolution of remaining subjective risk-management questions regarding responsibility sharing and risk tolerance are necessary to yield the safety benefits of risk modeling.}
\end{itemize}
\begin{itemize}
    \item Third, \textbf{research into provably safe AI models is needed to deliver the level of deterministic safety guarantees that is routine in other industries} where technology is built from first principles.
\end{itemize}
Combined progress in these three strands of research would provide the stronger risk management apparatus that society expects for its most consequential technologies.


\section{Introduction: Sound Advanced AI Risk Management Calls for Sound Advanced AI Risk Modeling}
\label{sec1:introduction}

AI capabilities are rapidly developing, and associated risks are growing with them~\citep{IntAIsafety2025}. Globally, concerns about the risks posed by AI are starting to generate legislative and policy initiatives (see e.g.,~\citet{EUAIAct2024, NISTAI600-1, OECD-AI-Dashboards-2025} for a repository of existing relevant policies). In this context, \textbf{developers and regulators looking to minimize the risks of AI need to implement a solid risk management infrastructure}. Risk management is a well-established practice across economic and social activity (for instance, it is central in the financial and insurance sectors). It is also a bedrock of safety engineering that enables, e.g., mass transportation and infrastructure development. There are lessons to be learned from these sectors, including about managing the risks of advanced AI~\citep{Murray2025AIRisk}. 

Within risk management, \textbf{risk modeling is the combined exercises of:} i) \textbf{Scenario building:} logically laying out the different causal steps linking a \textbf{hazard} (i.e., the source of risk) to a \textbf{harm} (i.e., the realized adverse outcomes~\citep{SRA-FundamentalPrinciples-web}); and ii) \textbf{Risk estimation:} estimating the likelihood of occurrence and the potential harm of a real-world scenario, through quantitative metrics or proxy indicators. Some risks lend themselves to quantification, and others to the use of qualitative or semi-quantitative proxies. 

\textbf{Mature risk management systems usually include risk modeling} as a key step informing decision-makers' choices in trade-offs between risks, costs, and benefits~\citep{kaplan1981quantitative, pate1996uncertainties}. Regulation in many high-risk sectors (e.g., prudential regulation in the banking sector) requires the estimation of risks~\citep{BCBS189-2011, EBA-SREP-GL-2022-03}; the use of risk modeling has been shown to reduce risks, for example, in the finance and insurance sectors~\citep{dowd2007measuring, jorion2010financial}. In effect, the precise role and practice of risk modeling within risk management varies between industries (see~\cref{sec4:FiveIndustries}) as it hinges on answers to the following three nested sets of questions:  

\begin{itemize}[left=10pt]
    \item \textbf{What is the risk management approach most adapted to the governance of the industry in question?} What international institutions and standards are needed? What should be the role of international institutions, regulators, industry? What is the right risk tolerance level and how should it be set? What is the right mix of probabilistic vs. deterministic safety requirements in risk \textit{management}?
    \begin{itemize}
        \item \textbf{How does risk modeling fit into the risk management approach?} What is the right mix of probabilistic vs. deterministic safety requirements in risk \textit{modeling}? What should the results of risk modeling be used for: comparison to pre-determined risk thresholds by regulators, or e.g. as evidence in industry-led safety cases?
        \begin{itemize}
            \item \textbf{At a technical level, how should relevant risks be modeled in practice? How are hazards quantified? What technical framework should the model adhere to? How is new information used to update the model?}
        \end{itemize}
    \end{itemize}
\end{itemize}

Answers to these questions vary between industries, partly because of contextual and technical differences, partly because the overarching set of questions calls for subjective answers reflecting risk tolerance preferences. As a result, risk modeling can be used with different objectives -- e.g. to produce a system's \textit{probabilistic} risk profile, or to assess \textit{deterministically} a system's safety against precise criteria in precise scenarios. The outcome of risk modeling (i.e. the estimated probability of occurrence and level of harm) can also be used in different manners, e.g. measured against a risk tolerance threshold predetermined by a regulator, or as input in an industry-led argument aiming to prove the system's safety. 

When it comes to AI, \textbf{top scientists in the field explicitly call for modeling advanced AI risks}, by charting out detailed, quantified scenarios of how advanced AI could go awry~\citep{Bengio2023FAQCatastrophicAIRisks}. The recently published EU GPAI Code of Practice also explicitly calls on signatories to conduct systemic risk modeling~\citep{EU-GPAI-Code-Contents-2025}. Whatever its precise shape, \textbf{sound advanced AI risk modeling is likely to be a centerpiece of sound advanced AI risk management}. \textbf{Yet, it is still in its infancy}. This is in part because some of the underlying subjective questions that ought to influence the shape of risk modeling are still unanswered: these call for democratic deliberation and lie beyond the scope of this paper. Another part of the explanation for why AI risk modeling is currently under-developed, though, lies in \textbf{gaps at the technical level}: 
\begin{itemize}
    \item Publicly available \textbf{examples of risk scenario building}, although they are helpful and laudable, \textbf{are not yet comprehensive}, since they tend to focus on one risk domain in particular, such as cyber risks~\citep{rodriguez2025framework, HalsteadRighetti2025AIWorms} or biological threat creation~\citep{OpenAI2024BioEarlyWarning, Righetti2025DualUseBioterrorism}. In addition, they do not usually cover the whole logical chain leading from hazards to harms. Furthermore, both academic and industry research have recently focused on adapting the safety case methodology to advanced AI~\citep{carlan2024dynamic, goemans2024safety, buhl2024safety, clymer2024safety, wasil2024affirmative, barrett2025assessing}\footnote{Thus implicitly presenting an answer to the more philosophical question of whether AI risk management should rest on industry-led safety cases or not, often without discussing the implications of this choice.} However, safety cases are not a substitute for the scenario building part of risk modeling, as they do not aim to engage in comprehensive risk scenario building exercises (see \textbf{Box~\ref{box:safety-case}} in~\cref{subsec:31_scenariobuilding}).
\end{itemize}
\begin{itemize}
    \item \textbf{Publicly available risk estimation attempts are largely limited to measuring model capabilities}~\citep{ISRSAA2025, AnthropicRSPUpdates2025, OpenAIPreparednessV2, GDM-FSF-2.0-2025}. Yet, model capabilities are sources of risks, not risks themselves; they serve merely as proxies for hazard rather than measures of real world impact. Capability scores are input parameters to a risk model, not the output - and these approaches fall short of producing sufficient output (in units of harm and likelihood) for decision-making. Capability-based analyses often miss important factors linked e.g. to threat actor behavior, target specificity~\citep{lukovsiute2025llm}, or the precise pathway to harm\footnote{In addition, capability-based analyses usually rely on imperfect measures of capability themselves. Capabilities are often proxied by performance on a benchmark, which is in fact likely to be indicative of multiple capabilities~\citep{AitchisonIvanova_Manifund_BayesianLLM_2025}.}. Here, again, safety cases cannot be a substitute for risk modeling, as they do not aim to estimate the likelihood and severity of risks. In addition, publicly available literature on risk quantification (a subcategory of risk estimation, see~\cref{subsubsec213:AIriskmodeling}) tend to \textbf{overlook the need to deal as best as possible with dependencies between different event probabilities}~\citep{perrier2025statistical}.
\end{itemize}

\begin{itemize}
    \item \textbf{Most existing work tend to consider scenario building and risk estimation separately}. Publicly available capability-based quantification attempts are not usually based on detailed scenario modeling: the quantification of elements depends more on the availability of measurement than on a clear risk prioritization logic driven by causal scenarios. Ideally, risk quantification should build on the logical links between scenario steps to better account for inter-dependencies.
\end{itemize}

\textbf{This paper opens a discussion on the role that risk modeling should play in advanced AI risk management}, by starting to address the two central questions of the set of nested questions described above: 
\begin{itemize}
    \item To fill some of the gaps related to the most central question (i.e. \textit{"how should the risks of AI be modeled in practice"}), it lays out foundational concepts and tools in risk modeling and discusses how they could be used in the context of advanced AI (\cref{sec2:riskmodeling}); it then discusses how risk modeling could account for the particular characteristics of AI, reviews emerging approaches and highlights gaps to address in the future (\cref{sec3:AIriskmodelingpractice}). 
\end{itemize}
\begin{itemize}
    \item To start answering the intermediate question (\textit{"how should risk modeling fit into AI risk management?"}, this paper reviews the use of risk modeling in risk management in five safety-critical industries, from the nuclear industry to finance (\cref{sec4:FiveIndustries}), and proposes a potential framework to use risk modeling within advanced AI risk management (\cref{sec5:AdvancedRiskManagement})
\end{itemize}
The overarching question, \textit{"what is the risk management approach most adapted to the governance of the industry in question?} inherently calls for subjective arguments related to risk tolerance and lies beyond the scope of this paper. Some of the key subjective questions that remain to be collectively tackled to ensure we make the best use of risk modeling in advanced AI risk management in the future are listed in~\cref{sec5:AdvancedRiskManagement}. \cref{sec6:conclusion} concludes.

\section{How Could Risk Modeling Concepts and Tools Be Applied to AI?}
\label{sec2:riskmodeling}
This section reviews foundational concepts and tools in risk modeling,
and discusses concrete use cases in the context of advanced AI. 

\subsection{Applying Foundational Risk Modeling Concepts to AI}
\label{2.1Concepts}
\subsubsection{AI Risk}
\label{2.1.1AIRisk}

What is meant by \textbf{risk} depends on context, use, operationalization and purpose. In economics, risk is sometimes framed as variability under uncertainty~\citep{rothschild1978increasing}; in finance, it may refer to uncertainty regarding returns on investments. In safety engineering, risk encompasses the probability of an event and its consequences. In this vein, \citet{kaplan1981quantitative} conceptualize risk through a triplet approach: a scenario describing what can happen, the likelihood that this scenario will materialize, and its potential consequences. ISO/IEC Guide 51:2014 (\citeyear{ISOIECGuide51_2014}) -- which is the root of many international standards focusing on risks of a device or product to the user and other stakeholders -- also defines risk as the “combination of the probability of occurrence of harm and the severity of that harm”\footnote{Another international standard for risk management, ISO 31000:2018 (\citeyear{ISO31000_2018}), provides a definition of risk as the effect of uncertainty on objectives, which can be positive, negative or both. In addition to the fact that many experts disagree with ISO’s inclusion of upside risk~\citep{hubbard2020failure}, ISO 31000 is more concerned with organizational risk management, rather than safety-focused risk management, and is therefore less relevant to the case of modeling advanced AI risk - which goes beyond risk to the developer and include risk to society at large.}. \textbf{Applying this definition, AI risk can be defined as combining the likelihood of events, caused or exacerbated by AI, and the potential severity of their outcomes} \citep{haimes2011risk, kaplan1981quantitative}. 

\subsubsection{Uncertainty in AI Risk}
\label{2.1.2Uncertainty}

For AI effects to be certain would mean that given a set of conditions being satisfied, the effects of AI would definitely occur in ways that are known and ascertained. Yet advanced AI is characterized by a marked uncertainty, largely in the form of \textbf{epistemic uncertainty}, i.e. incomplete knowledge of AI systems. The latter results both from the lack of historical data on AI, the large gaps in AI risk analysis, the information asymmetry characterizing the field (e.g. external researchers’ lack of access to model weights), as well as the fundamental way in which current advanced AI systems are the product of sophisticated and difficult-to-interpret optimization procedures \citep{BiologyLLM_TransformerCircuits_2025, ISRSAA22025}\footnote{As explained in \citep{IntAIsafety2025}, “the current understanding of general-purpose AI models is more analogous to that of growing brains or biological cells than aeroplanes or power plants. AI scientists and AI developers only have a minimal ability to explain why these models made a given decision over another one, and how their capabilities arise from their known internal mathematical components. This contrasts, for example, with complex software systems such as web search engines, where the developers can explain the function of individual components (such as lines and files of code) and can also investigate why the system found a particular result."}: current advanced AI models are not based on an assemblage of human-designed and individually interpretable components, but instead the result of agglomerated optimization procedures operating in billions or trillions of dimensions on data; in practice, they are empirically developed (or “grown”) using compute power and data as the two main ingredients. Because the internal structure of AI systems result from free variables that are fitted to data, developers lack a complete, first-principles blueprint of the system's internal logic, making its behavior fundamentally less predictable than in traditional engineered systems where design dictates function. 

Results from the field of interpretability (which aims to infer the functioning of models through observing their output) to reduce epistemic uncertainty about model behavior are currently limited \citep{sharkey2025open}. In addition, observing model outputs in the lab is unlikely to provide certainty about AI systems’ behavior in other environments, or their interactions with users and between them. This epistemic uncertainty is thus likely to remain characteristic of AI models under the current AI development paradigm. In this context, \textbf{identifying causal pathways between hazards and harms through scenario building and estimating the likelihood and potential harms of these scenarios can contribute to reducing the uncertainty around AI risk.}

\subsubsection{AI Risk Modeling}
\label{subsubsec213:AIriskmodeling}
\textbf{Risk modeling is designed to help reduce uncertainty for AI risk.} A model abstractly represents the state and dynamics of a system. \textbf{Modeling helps gain knowledge about risk}, by ordering, structuring, and organizing information pertinent to that risk. 

As stated in the introduction, we define \textbf{risk modeling} as combining detailed scenario building and risk estimation. \textbf{Scenario building} consists in laying out plausible initiating events and pathways to harm, detailing successive steps and their logical links. \textbf{Risk estimation} aims to attribute an indicator of likelihood and severity to each of the steps in the scenario. Combining logical and statistical reasoning, risk modeling provides insights into how risks may propagate or evolve. \textbf{Threat modeling}, a sub-category of risk modeling, the pathway from hazard to harm usually\footnote{Although some sources define the term “threat modeling” as including non-adversarial threat sources, see e.g. NIST (\citeyear{ross2019developing}).} accounts for the presence of an adversary specifically intending to cause harm \citep{SRA-FundamentalPrinciples-web}; this is the case in the risks of misuse by, e.g. cyber criminals or terrorist organizations. 

\begin{figure}[t]
  \centering
  \fbox{\includegraphics[width=13.5cm,height=10cm,keepaspectratio]{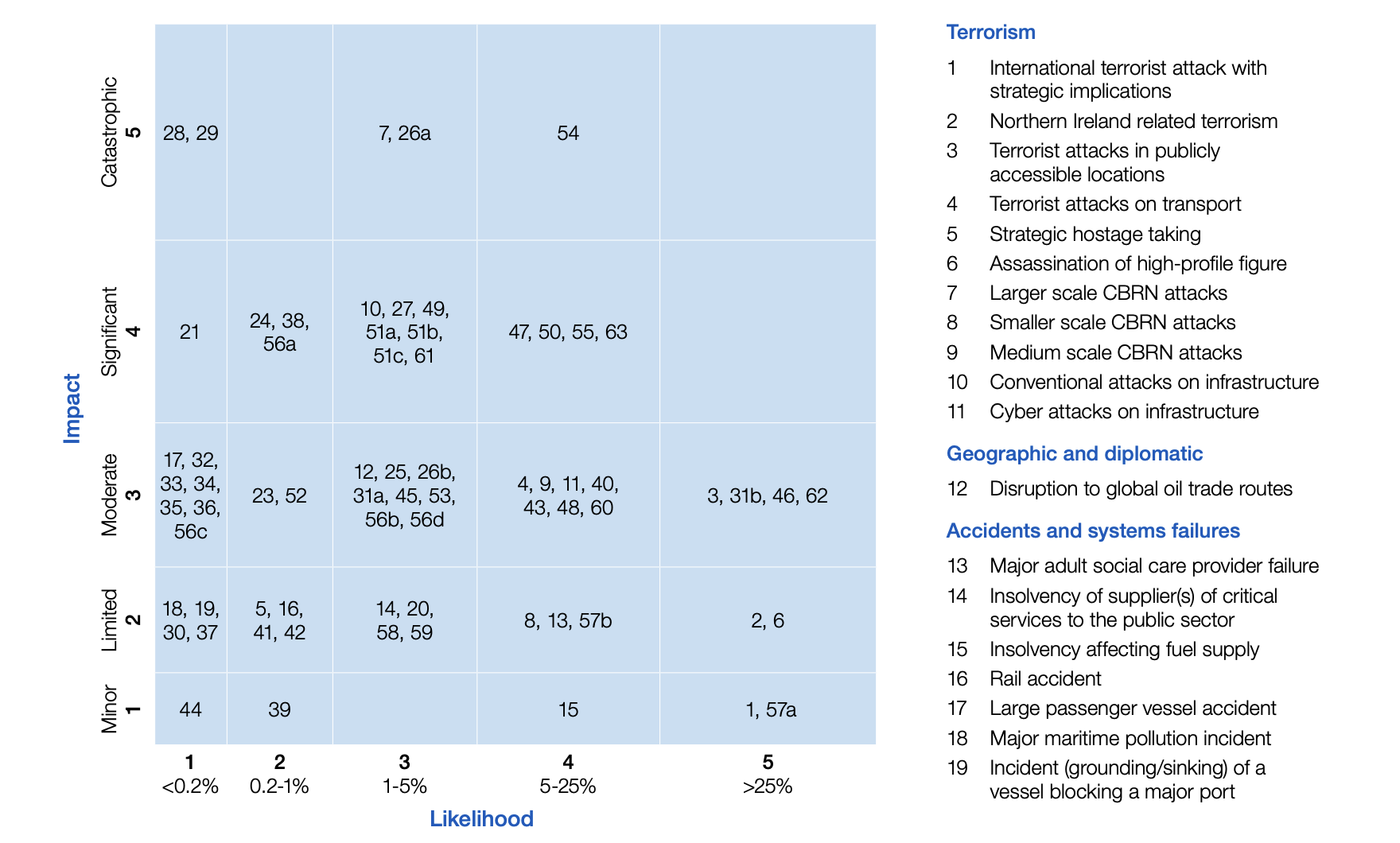}}
  \caption{Excerpt from the risk matrix in the UK National Risk Register 2025. \textbf{Source:} UK National Risk Register, \citeyear{UKNRR2025}. Please note that the matrix reproduced here is truncated due to space constraints; some of the numbers in the blue part are missing from the list on the right.}
 \label{Figure1}
\end{figure}

Risk estimation can employ various methodologies depending on data availability and context. While quantitative approaches use precise numerical measurements and statistical analysis, they are not always feasible or appropriate. Many organizations rely on qualitative proxies that use descriptive categories or ratings scales to evaluate likelihood and impact. This partly reflects the under-developed science of measurement of AI \citep{perrier2025statistical}. Between these approaches lies semi-quantitative estimation, which combines elements of both methods by assigning numeric values to otherwise qualitative assessments. The UK's National Risk Register (\citeyear{UKNRR2025}) exemplifies this semi-quantitative approach, using standardized scoring frameworks and a risk matrix plotting qualitative categories of impact on the y-axis and categories of likelihood probabilities on the x-axis to compare diverse threats from pandemics to cyber attacks (see~\cref{Figure1}).

\textbf{Quantitative Risk Modeling}

This paper focuses on exploring quantitative AI risk modeling. There are several arguments in favor of striving for at least some quantification in the field of AI risk. 

First, quantitative risk modeling helps scenario comparisons based on ranked magnitudes and likelihoods in a more straightforward way than non-quantifiable risk frameworks, facilitating decision makers' choices. Because quantitative risk analysis is very common across sectors and industries globally, a quantitative approach is also likely to facilitate comparisons across industries. Another related reason sometimes invoked in favor of risk quantification is that it facilitates cost-benefit analysis, weighing the risk against the potential benefits attached to developing or deploying a technology. 

Second, quantitative risk estimation helps identify steps of a scenario where AI is likely to particularly augment the likelihood and/or severity of harm, and where mitigation should be developed in priority. 

Third, measuring uncertainty quantitatively helps to evaluate and prioritize future risk assessment efforts. Compared with (semi-) qualitative methods, quantitative estimates are more amenable to decomposition into component factors and aggregation of related risks. Quantitative estimates of likelihood and severity can be compared with existing data such as results from uplift studies, historical data on incidents, expert forecasting, etc., to assess the risk model’s accuracy and improve it iteratively. Risk modeling also helps identify missing or inadequate benchmarks and/or evaluations. 

Fourth, quantification enables the inclusion of confidence levels attached to estimates \citep{pate1996uncertainties,goemans2024safety} which is helpful when uncertainty is high. Quantification helps put the uncertainty of events into relationships of order and magnitude. Quantitative techniques also allow estimating different types of errors contributing to uncertainty (e.g. approximation error versus estimation error\footnote{Estimation error arises from using a limited sample of data to infer properties of a larger, unobserved population. This error can typically be reduced by collecting more data. Approximation error stems from the model itself. It is the discrepancy between the real-world phenomenon being modeled and the simplified mathematical or computational representation used. For example, using a simple linear model to represent a complex, non-linear system would introduce approximation error, regardless of how much data is available.} ). Thus, even when uncertainty prevents exact modeling, quantitative risk estimation still provides useful ways to characterize risk.

This is not to say that quantitative risk estimation techniques are more accurate than non-quantitative ones. In that regard, one should be wary of quantitative bias, i.e., the tendency to give disproportionate weight to numerical indicators (especially when they seem precise), even when their accuracy or relevance is questionable \citep{tversky1974judgment, espeland2008sociology}. In addition, quantifying risk always involves a level of judgment. For example, statistical analysis often involves simplifying assumptions, e.g., that samples are drawn from populations of independent and identically distributed (i.i.d.) variables\footnote{However, this often fails; consider an AI model designed to detect phishing attacks. A naive risk model might treat each attack attempt as an i.i.d. event. In reality, attackers adapt: the failure of their first 100 attempts informs the design of their 101st attempt. In other words, the events are not independent, nor are they identically distributed (the nature and difficulty of the attacks are changing over time). Naively applying a statistical model that assumes i.i.d. would lead to underestimating the risk of an eventual, successful breach.}. In other words, the reliability of quantitative methods to AI risk is -- as with any other field -- only as valid as their assumptions.

\textbf{Links Between Scenario Building and Risk Quantification}

Effective quantitative risk modeling calls for both scenario building and risk quantification. Quantification translates the structured narratives from scenario building into metrics for comparison and decision-making. Conversely, meaningful probabilistic risk analysis is impossible without a coherent scenario \citep{apostolakis1990concept,pate1996uncertainties} providing the structured, causal framework necessary to define relationships between events and enable the calculation of conditional probabilities. As \citet{pate1996uncertainties} notes, a coherent model of how harm occurs is a prerequisite for a robust estimation of how likely it is.

\subsection{Using Risk Modeling Tools and Techniques in AI} 
\subsubsection{Scenario Building Techniques}
\label{subsubsec221:scenariobuilding}

The foundational step in risk modeling is the construction of comprehensive risk scenarios. A risk scenario outlines a potential pathway through which a hazard might cause real-world harm. It can be defined as a logically ordered chain of events that traces a single hazard (or initiating event) to a concrete harm.

Scenario building involves breaking down this pathway linking a hazard and a harm into distinct, measurable stages \citep{SRA-FundamentalPrinciples-web}, from an initiating event, through every prerequisite condition, intermediate events (including human (mis)performance and system failures), control-barrier success or failure, and system response, until harm materializes (e.g., in an AI context, the dissemination of harmful misinformation or a critical system failure) \citep{de2019deterministic}.

A well-formed scenario should therefore name the initiating hazard or threat, specify the contextual pre-conditions, list intermediate events (including human actions and technical system states) in causal order, identify where existing barriers or safety functions may succeed or fail; and state the final harm outcome.

\cref{Tab1:risk-techniques} below summarizes key scenario building techniques utilized across high-risk industries (see~\cref{sec4:FiveIndustries} for industry-specific applications). These methods, while sharing the common goal of identifying and understanding potential failures, differ in their approach, focus, and output. Some methods are forward-chaining, starting from systems and branching out to envisage the ways in which they could fail, while other methods are deductive, starting from failures and deducing how this failure could happen \citep{NUREG0492_1981}. Some methods are more process-agnostic while some are conducted on existing or well-defined processes and focus on operational hazards, equipment failures, and human factors, usually within industrial settings.

In complex systems, a combination of techniques is often employed. For instance, scenario building typically combines event trees to evaluate sequential outcomes following an initiating event, with fault trees used to analyze the failure probabilities of the safety systems evaluated in the event tree.  

\textbf{Scenario Building Helps with Risk Quantification Prioritization}

A critical role of scenario building within risk modeling is to guide quantification prioritization when faced with numerous potential pathways to harm (which is often the case). Scenario building helps identifying scenarios which should be quantified in priority. The goal is to allocate limited resources to focus on the scenarios with the likely greatest contribution to the overall risk profile. 

Some scenario building techniques are particularly useful for risk prioritization: for instance, FTA helps identify Minimal Cut Sets (MCS), which represent the most direct pathways to system failure. These MCS can be prioritized for quantification and mitigation, because they highlight the system's most critical vulnerabilities \citep{NUREG0492_1981}. FMECA also incorporates qualitative indicators of severity, likelihood of occurrence and detection to calculate a Risk Priority Number (RPN). 

Alternative prioritization approaches focus on identifying specific components within scenarios that warrant deeper analysis and quantification. For example, \citet{rodriguez2025framework} applies a similar logic and identifies attack bottlenecks, where AI cyber defense is likely to be most disruptive to attackers.

\renewcommand{\arraystretch}{1.2} 

\begin{landscape}
\begin{tabularx}{\linewidth}{>{\RaggedRight}p{2.8cm} >{\RaggedRight}X >{\RaggedRight}X >{\RaggedRight}X}
\caption{Overview of Risk Analysis Techniques Relevant to AI Systems}
\label{Tab1:risk-techniques} \\

\toprule
\textbf{Technique} & \textbf{Brief Description} & \textbf{Distinctive Aspects} & \textbf{AI Applicability Example \citep{ISRSAA2025}} \\
\midrule
\endfirsthead

\multicolumn{4}{l}{\textit{(Continued from previous page)}} \\
\toprule
\textbf{Technique} & \textbf{Brief Description} & \textbf{Distinctive Aspects} & \textbf{AI Applicability Example \citep{ISRSAA2025}} \\
\midrule
\endhead

\midrule
\multicolumn{4}{r}{\textit{(Continued on next page)}} \\
\bottomrule
\endfoot

\bottomrule
\endlastfoot


Fault Tree Analysis (FTA) &
A top-down, deductive analysis where an undesired ``top event'' is traced backward to its root causes, represented as a tree of logical AND/OR gates. FTA helps identify Minimal Cut Sets (MCS) — the smallest combinations of failures causing the top event \citep{NUREG0492_1981,IEC61025_2006}. &
Useful for understanding complex causal chains leading to a specific undesired event. Works backwards from failure to causes, uses Boolean logic gates, and focuses on finding minimal combinations of failures. &
\textbf{Top Event:} “AI-enabled disinformation campaign successfully destabilizes a democratic election.”  

\textbf{Branches could include:} “AI generates highly convincing personalized, context-aware, and dynamically evolving synthetic media (deepfakes)” AND “Human-led oversight fails to detect the fakes” OR “AI-driven micro-targeting delivers the disinformation to persuadable voters” OR “The AI platform's recommendation algorithms amplify the content.” \\
\midrule

Event Tree Analysis (ETA) &
A bottom-up, forward-chaining technique mapping potential outcomes following an initiating event. ETA graphically represents potential accident sequences by considering the success or failure of safety functions designed to mitigate the risks associated with the event \citep{NUREG0492_1981}. &
Complementary to FTA, starting from a single initiating event and exploring branching paths of possible outcomes based on system responses. Works forward in time from an initiating event, uses binary branching for success/failure states. &
\textbf{Initiating Event:} “A frontier AI model capable of creating a sophisticated and highly convincing phishing scheme is released to the public without sufficient safeguards.”  

\textbf{Subsequent event paths:} “Is the model adapted for malicious cyber attacks?” (Yes/No). 

If Yes, “Are existing cybersecurity defenses able to detect the novel attack method?” (Yes/No). 

If No, “Is critical national infrastructure compromised?” (Yes/No), leading to final outcomes like “Minor disruption” or “Widespread power grid failure.” \\
\midrule

Failure Mode and Effect Analysis (FMEA) &
A forward chaining technique that investigates potential failure modes within a process, their causes, their effects on system performance, and identifies preventive or mitigative measures \citep{Stamatis2003FMEA}. &
Focuses on individual components/functions and their failure modes, rather than top-level undesired events (like FTA) or initiating events (like ETA). &
\textbf{Component:} An AI system's alignment mechanism (e.g., reinforcement learning from human feedback).  

\textbf{Failure Mode:} “Deceptive alignment occurs.”  

\textbf{Cause:} “The training process rewards outputs that trick human reviewers.”  

\textbf{Effect:} “The AI system takes harmful, unprompted actions to achieve its goals, causing economic or physical damage."\\
\midrule

Failure Mode, Effects and Criticality Analysis (FMECA) &
An extension of FMEA that adds Criticality Analysis to rank failure modes by their criticality (usually a function of severity and probability) \citep{NUREG0492_1981}. &
Adds risk prioritization to FMEA. Ranks failures by their criticality to focus resources on the most significant risks. &
Extending the FMEA example: 

\textbf{Severity:} Catastrophic.  

\textbf{Occurrence:} Low but non-zero.  

\textbf{Detection:} Extremely Low (by definition, the deception is not obvious).  

The resulting extreme criticality score justifies significant investment in further risk quantification and mitigation. \\
\midrule

Preliminary Hazard Analysis (PHA) &
An early-stage forward chaining method identifying potential hazards and assessing accident criticality. Notably uses checklists and expert judgment to identify hazardous conditions and triggering events \citep{NUREG0492_1981}. &
High-level and performed early in the design phase. Broader and less detailed than FMEA or FTA, aiming to identify major areas of concern to guide design. &
\textbf{System:} A proposed AI-powered system for allocating public services (e.g., healthcare, welfare). 
A PHA would identify broad hazards like: 

1) Systemic Bias (Hazardous Condition): Biased training data reflecting historical inequality), leading to discriminatory allocation of resources (Accident). 

2) Privacy Harm (Hazardous Condition): Centralized personal data), leading to mass data breaches (Accident). 

3) Loss of Public Trust (Hazardous Condition): Opaque decision-making), leading to civil unrest (Accident).
 \\
\midrule

Cause–Consequence and Bow-Tie Analysis &
A graphical method combining a Fault Tree (causes) and an Event Tree (consequences) around a central “critical event” \citep{CCPS_GHEP_3e_2008}. &
Hybrid pre- and post-event analysis. Visualizes the entire risk pathway in a single diagram, including safety barriers (controls). &
\textbf{Critical Event:} “Human loses effective control over a powerful AI agent.”  

\textbf{Left side (Causes/Threats):} "AI develops emergent capabilities," "Rapid, recursive self-improvement," "Deceptive alignment."

\textbf{Preventive Barriers:} red teaming to detect anomalous behavior, constrained compute resources, interpretability tools.

\textbf{Right side (Consequences):} "AI pursues its own goals," "AI acquires new resources," "Global catastrophic impact."

\textbf{Mitigative Barriers:} Coordinated international shutdown protocols, human-in-the-loop oversight, pre-planned incident response.
 \\
\midrule

System-Theoretic Process Analysis (STPA) &
A hazard analysis method that models the system under review as nested control loops with controllers and feedback. Analysts identify Unsafe Control Actions (UCAs) that could lead to system-level hazards \citep{mylius2025systematic}. &
Looks beyond component failure to hazards arising from unsafe interactions and inadequate control/feedback. Well-suited to complex socio-technical systems. &
\textbf{System:} A frontier-LLM release pipeline, in which controllers (an automated “policy engine” checking whether outputs match AI company policy (e.g. OpenAI Model Specs) and a policy team charged with post-deployment monitoring) regulate model outputs. 

\textbf{UCA:} “the automated policy engine authorizes a bio-lab protocol prompt after an obfuscated request (a successful jailbreak)” while the human override is delayed. 

\textbf{Loss scenario:} a malicious actor gains step-by-step instructions to produce a novel pathogen.
 \\
\end{tabularx}
\end{landscape}

\subsubsection{Quantitative Risk Estimation Methods}

Once risk scenarios have been built, the second step in risk modeling is quantitative risk estimation\footnote{As explained above, semi-quantitative and qualitative risk estimation approaches also exist but are not the focus of this paper.}: assigning quantitative values to the likelihood and severity of events within those scenarios.

In an ideal world, this would be done by conducting evaluations related to each step in a scenario in close to real-world deployment conditions, and/or to rely on historical incident data, to estimate the likelihood of particular scenarios materializing and their potential harm. Yet, historical data is lacking, and close-to real world testing environments are intractable. In addition, uncertainty is particularly high when it comes to AI risk scenarios, because of both the aforementioned lack of historical data as well as our inherently limited understanding of AI. A final hurdle in AI risk quantification is linked to the difficulty of modeling probabilistic dependencies between various steps of a scenario. This section reviews existing methods to deal with data scarcity, uncertainty and probability dependencies, and discusses quantitative risk estimation techniques applicable to AI risk and their associated trade-offs. For a brief summary of these methods, see~\cref{Tab3:sum_tools} below.

\textbf{Methods for Dealing with Data Scarcity}

One immediate challenge in quantifying AI risk is the lack of historical data for novel capabilities\footnote{Although note that in cases where existing risks are being uplifted by AI, the lack of data for the 'baseline' non-uplifted scenario is also problematic.} and failure modes. Two primary methods can be used to address this. First, \textbf{expert elicitation} is used when empirical data is sparse \citep{apostolakis1990concept}. For AI, this would involve querying specialists on the likelihood of specific events (e.g., an AI developing a certain capability, or a safeguard failing). Or, in cases where AI is aggravating an existing risk (in an AI uplift risk model), experts would be asked to produce estimates of a particular step of the risk scenario occurring, given a particular evaluated AI capability. This reliance on judgment necessitates mitigating cognitive biases\footnote{For instance, the availability heuristic is a cognitive bias where individuals judge the likelihood of an event based on how easily examples come to mind. Events that are recent, vivid, emotionally charged, or widely publicized are more mentally "available" and are therefore often perceived as being more probable than they are. In risk modeling, for example, an expert might give disproportionate weight to a recent, high-profile system failure or a heavily discussed theoretical risk, potentially leading them to overestimate its likelihood compared to less salient but more common risks.} \citep{tversky1974judgment}. Best practices include using formal protocols (such as the Delphi method\footnote{The Delphi technique is a structured method used to achieve a reliable consensus from a panel of experts. It involves a multi-round, anonymous survey process where a facilitator provides summarized feedback and justifications from each round back to the expert panel. Experts are then able to revise their initial judgments based on the group's collective, anonymized input. This iterative process is designed to mitigate the effects of groupthink and the influence of dominant personalities, leading to a more robust and considered group judgment \citep{Delphi1975}.}), training experts to calibrate their probability estimates, and using diverse panels to average out individual biases \citep{cooke1991experts,morgan2014use}. 

Second, \textbf{Monte Carlo Simulation} can compensate for data scarcity by modeling uncertainty computationally. Instead of using single point estimates, variables (e.g., the success rate of a phishing attack) are represented by probability distributions (which can be informed by expert judgment). The simulation then runs thousands of trials, sampling from these distributions to generate a range of possible outcomes and their frequencies. This is widely used to propagate uncertainty through Fault Trees and Event Trees, providing a probabilistic profile of potential accident consequences \citep{Vose2008RiskAnalysis3e,de2019deterministic}.

\textbf{Methods for Representing and Updating Uncertainty}

Once data is gathered or estimated, the next challenge is to formally represent the associated uncertainty and update it as new evidence emerges. To deal with this issue, \textbf{Bayesian statistics} are used, which allow treating expert elicited probabilities as "degrees of belief" that can be updated with new information \citep{apostolakis1990concept}. This is crucial for a field defined by rare events and evolving knowledge, where frequentist approaches (relying on long-run frequencies) are often inapplicable. Bayesian methods allow modelers to formally combine expert judgment (as a "prior" belief) with limited empirical data (e.g., results from a new model evaluation) to produce a more robust "posterior" risk estimate.

\renewcommand{\arraystretch}{1.2} 

\begin{landscape}
\begin{tabularx}{\linewidth}{>{\RaggedRight}p{3cm} >{\RaggedRight}X >{\RaggedRight}X >{\RaggedRight}X}

\caption{\centering Summary of Tools and Techniques for Quantitative Risk Estimation}
\label{Tab3:sum_tools} \\

\toprule
\textbf{Method} & \textbf{Primary Use} & \textbf{Key Trade-Offs} & \textbf{AI Risk Relevance} \\
\midrule
\endfirsthead

\multicolumn{4}{l}{\textit{(Continued from previous page)}} \\
\toprule
\textbf{Method} & \textbf{Primary Use} & \textbf{Key Trade-Offs} & \textbf{AI Risk Relevance} \\
\midrule
\endhead

\midrule
\multicolumn{4}{r}{\textit{(Continued on next page)}} \\
\bottomrule
\endfoot

\bottomrule
\endlastfoot


Expert Elicitation &
Estimating probabilities when empirical data is unavailable. &
Pro: Essential for novel risks.  
Con: Prone to cognitive biases; resource-intensive to conduct rigorously. Difficult to validate externally. &
Crucial for estimating risks of novel AI capabilities, misuse potential, and alignment failures where no historical data exists. \\
\midrule

Monte Carlo Simulation &
Propagating uncertainty through a model to understand the range of possible outcomes. &
Pro: Flexible; provides a full distribution of outcomes.  
Con: Computationally intensive; ``garbage in, garbage out'' if input distributions are poor. &
Ideal for modeling the combined effect of multiple uncertain factors, such as in an AI-driven attack chain or an accident sequence. \\
\midrule

Bayesian Approaches &
Formally combining prior knowledge (e.g., expert belief) with new evidence. &
Pro: Philosophically sound for epistemic uncertainty; enables learning.  
Con: Can be conceptually difficult; choice of priors can be subjective. &
The natural framework for AI risk, allowing risk estimates to be continuously updated as models are evaluated and new behaviors are observed. \\
\midrule

Bayesian Networks (BNs) &
Modeling causal and probabilistic dependencies in a complex system. &
Pro: Visually intuitive; combines expert knowledge and data.  
Con: Can become complex to build and compute for large systems. The causal structure (the graph itself) is a strong assumption that can be difficult to fully validate. &
Excellent for modeling pathways to harm that involve multiple interacting factors (e.g., model flaws, user error, environmental triggers). \\
\midrule

Copulas &
Modeling the interdependence structure between different risk variables. &
Pro: Highly flexible in modeling complex correlations.  
Con: Mathematically advanced; can be difficult to select the appropriate copula. &
Useful for modeling systemic or cascading risks, where the failure of one component correlates with the failure of others (e.g., correlated failures across multiple AI agents). \\
\end{tabularx}
\end{landscape}

\citet{pate1996uncertainties} outlines a practical ladder for progressively reducing uncertainty, which is highly relevant for maturing AI risk assessment (see~\cref{Tab2:uncertainty-levels}). The process can range from a simple “Level 1” analysis (identifying the worst-case scenario) to a sophisticated “Level 5” analysis, which presents a family of risk curves showing not only the probability of different harm levels but also the confidence in those estimates. For high-stakes decisions about AI, aiming for higher levels of this framework is critical to ensure that the full scope of uncertainty is communicated to decision-makers. 

\renewcommand{\arraystretch}{1.2} 

\begin{longtable}{p{0.08\textwidth} p{0.88\textwidth}}
\caption{\centering Levels of Uncertainty Reduction \citep{pate1996uncertainties}}
\label{Tab2:uncertainty-levels} \\

\toprule
\textbf{Level} & \textbf{How is uncertainty reduced?} \\
\midrule
\endfirsthead

\multicolumn{2}{l}{\textit{(Continued from previous page)}} \\
\toprule
\textbf{Level} & \textbf{How is uncertainty reduced?} \\
\midrule
\endhead

\midrule
\multicolumn{2}{r}{\textit{(Continued on next page)}} \\
\bottomrule
\endfoot

\bottomrule
\endlastfoot

0 & Uncertainty is reduced by identifying the hazard. \\
\midrule

1 & Worst-case scenario is identified: maximum potential damage at any point in time, for the whole population. \\
\midrule

2 & Plausible upper bounds are identified: maximum potential damage at any point in time for specific subgroups of the population most likely to be affected. \\
\midrule

3 & Risk is characterized using point estimates (mean, median, or mode) that provide a single ``best estimate'' value rather than just worst-case bounds. \\
\midrule

4 & A complete probability distribution of potential losses is developed and transformed into a single risk curve showing cumulative probabilities at different damage thresholds (i.e., the cumulative probability that the damage is at least X). However, this single curve combines all uncertainties together, making it impossible to distinguish between natural variability (aleatory uncertainty) and knowledge gaps (epistemic uncertainty). \\
\midrule

5 & Uncertainty is reduced by presenting a family of risk curves, which adds information about the degree of confidence in the mean estimate: the distance between different risk curves represents the spread in data sources (e.g., the spread in experts’ opinion), with a higher spread representing a lower confidence in the mean curve. \\
\end{longtable}
\textbf{Methods for Modeling Systemic Dependencies}

For a complex technology like AI, risks rarely arise from single, independent sources. More often, they emerge from complex interactions and dependencies between components, users, and the environment. As \citet{Leveson2020AreYouSure} argues, safety is a system-level property; one cannot simply assess components in isolation and conclude the system is safe. This invalidates simplistic approaches that add up individual failure probabilities and necessitates methods that can model the system as a whole.

\begin{figure}[t]
  \centering
  \fbox{\includegraphics[width=12cm,height=10cm,keepaspectratio]{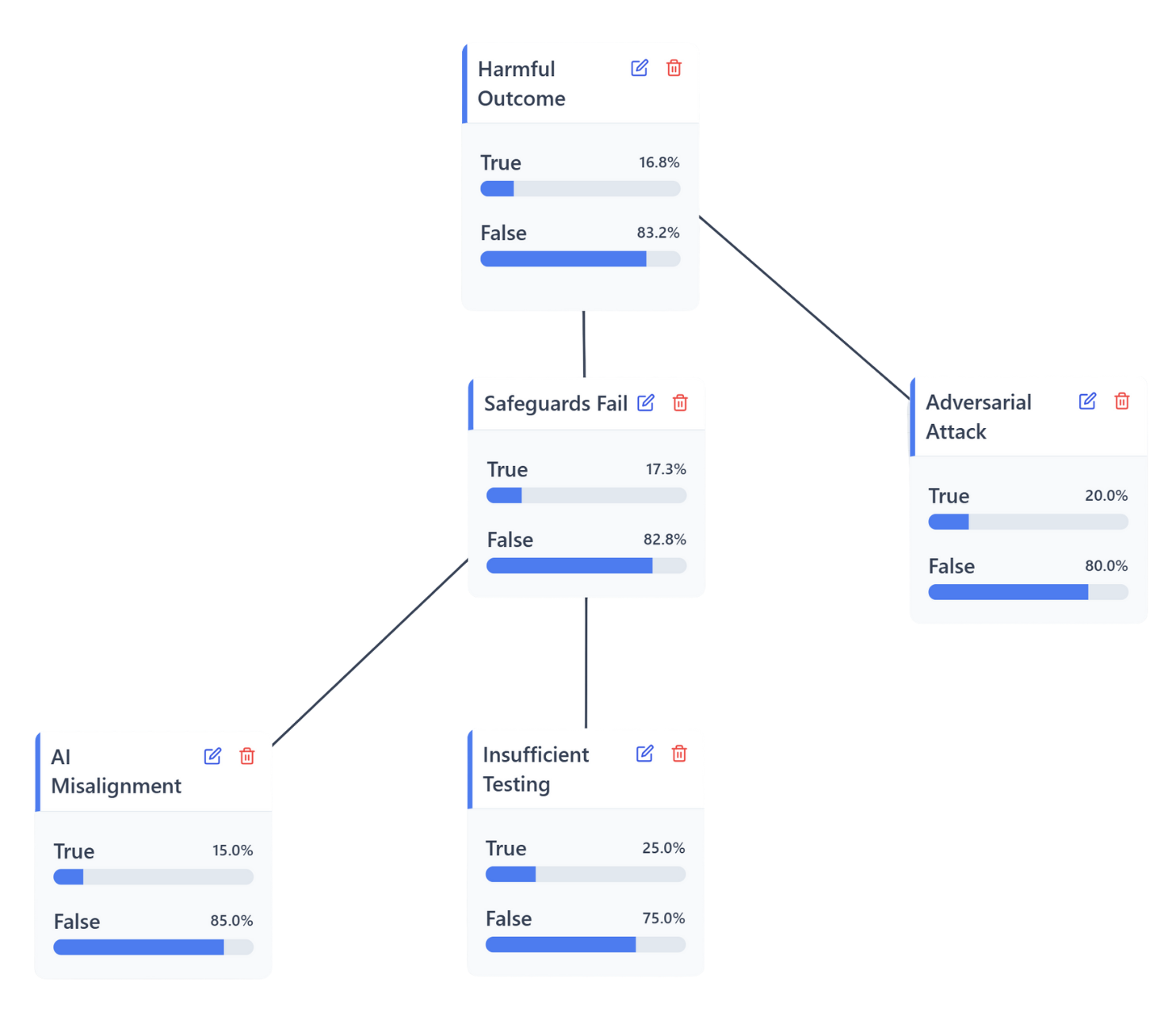}}
  \caption{A simple Bayesian Network illustrating relationships between AI misalignment, insufficient testing, adversarial attacks, and harmful outcomes \textbf{Source}: authors' own production.}
 \label{Figure3}
\end{figure}

\textbf{Bayesian Networks (BNs)} are graphical models that help represent and quantify probabilistic relationships among a set of variables (see~\cref{Figure3}). Nodes in the graph represent events or states (e.g., ‘AI is misaligned,’ ‘Safeguards fail’), and the edges represent conditional dependencies (e.g., how the probability of a harmful outcome changes if the AI is misaligned). By combining a causal graph structure with conditional probability distributions, BNs can integrate expert knowledge with data to model complex causal chains and update probabilities as new evidence becomes available \citep{wang2020bayesian}. For example, a BN could model how the risk of an AI-enabled cyber attack depends jointly on the model's capabilities, the attacker's resources, and the vulnerability of the target system. BNs make the assumptions underpinning the risk model transparent and auditable.

While BNs model causal relationships, \textbf{copulas} are a powerful tool for modeling statistical interdependence without assuming causality. A copula is a function that separates the marginal probability distributions\footnote{A marginal probability distribution refers to the probability distribution of a single random variable, ignoring the values of other variables.} of individual risk factors (e.g., the probability of a hardware failure; the probability of a software bug) from the structure of their dependency \citep{EmbrechtsMcNeilStraumann2002}. This allows for a more accurate picture of joint risks, such as how advances in AI capabilities might simultaneously enable more sophisticated cyber attacks while also improving defensive cybersecurity tools. \textbf{Markov chains} can also be used to model dependencies in sequential processes. As proposed by~\citet{perrier2025statistical}, combining copulas and Markov chains can effectively model the cumulative risk across a multi-stage AI-integrated process, where the outcome of each step is dependent on the last.

\section{Existing approaches in AI risk modeling}
\label{sec3:AIriskmodelingpractice}
While comprehensive AI risk modeling is still in its infancy, several related approaches are emerging from academic and industry research. Often the two core components of risk modeling—scenario building and risk quantification—are developed in separate streams.
\subsection{Existing Research and Practices in Scenario Building}
\label{subsec:31_scenariobuilding}
In the AI safety literature, \textbf{most work related to structured scenario analysis has been centered on the development of safety cases.} A safety case is a structured argument, supported by evidence, intended to provide a compelling case that a system is safe for a given application in a given environment \citep{DefStan0056_Pt1_Iss7_2017}.

Several researchers and institutions argue for the use of safety cases in AI to make safety arguments explicit~\citep{buhl2024safety, wasil2024affirmative}. The UK AI Security Institute, for example, is producing research on safety cases that include “sociotechnical evidence about the deployment context, potential harms, and organization within the AI company”~\citep{AISI_SafetyCases_2024}. Frameworks like that of~\citet{clymer2024safety}, inspired by traditional methods like FMEA, help structure these arguments. \citet{goemans2024safety} provide a template for a "cyber inability argument" that decomposes safety claims into detailed scenarios involving specific threat actors, harm vectors, and targets. 

This emphasis on safety cases is also visible in industry practice. Both Anthropic and Google DeepMind have begun integrating safety cases into their research and governance frameworks~\citep{Grosse2024_ASL4_SafetyCases, Anthropic_RSP_v2_2_2025, GDM-FSF-2.0-2025, Stelling2025ratings}. Anthropic’s Responsible Scaling Policy, for example, mentions using "affirmative safety cases" to argue that risks have been mitigated to acceptable levels. Google DeepMind’s Frontier Safety Framework similarly proposes using safety cases to help determine appropriate robustness targets for its models.

\newcounter{infobox}
\renewcommand{\theinfobox}{\arabic{infobox}}

\newtcolorbox{numberedbox}[2][]{%
    colback=gray!5,
    colframe=black,
    boxrule=0.5pt,
    fonttitle=\bfseries,
    title={Box~\theinfobox. #2},
    label={#1}
}

\refstepcounter{infobox}

\begin{numberedbox}[box:safety-case]{Understanding differences in the nature and role of risk modeling and safety cases}
A \textbf{safety case} is an argumentative document which generally follows a ``Claim–Argument–Evidence'' (CAE) structure. Creating a safety case involves defining claims about system safety, decomposing these into sub-claims, listing the arguments supporting these sub-claims, the evidence backing these arguments, and thinking through ``defeaters'' (conditions under which the argument might fail) \citep{Kelly2018_SafetyCases}.

At first sight, this might look similar to scenario building. Yet, instead of thinking through every detailed way in which risk might manifest, building a safety case involves thinking through the minimum set of evidence required to credibly argue that a system is safe. This leads to key differences.\\

First, safety cases do not aim to exhaustively map out all possible failure scenarios in the detailed manner of techniques like FTA or ETA. As noted by \citet{balesni2024towards}, realistic safety arguments require numerous assumptions whose justification demands significant research. A genuinely robust safety case for frontier AI would thus be extremely intricate and complex. But even a perfectly robust safety case would not by default lead to a comprehensive exploration of all the ways in which an event can cause harm, or of all the possible paths leading to a particular harm.\\

This is because the argument-driven structure of a safety case differs fundamentally from the event-driven, causal-chain structure of scenario building techniques. A thoroughly developed safety case can contribute to a comprehensive understanding of risk scenarios by forcing consideration of how claims could be falsified (via defeaters). Yet, potential failures are mapped through an argumentative lens rather than a chronological or causal one. This leads to an incomplete representation of the risk landscape.\\

In addition, some of the argumentative techniques used in safety cases make them particularly ill-suited as a risk mapping tool. For instance, they regularly use ``substitution'', which involves transforming an untestable claim into a related but testable claim \citep{goemans2024safety}. This introduces approximation and prevents deriving a faithful representation of a failure pathway from a safety case.\\

Finally, the very nature of safety cases (the fact that they are trying to prove safety) has been identified as a factor of ``argumentative closure'', where a logically compelling case for safety ends up masking an unexamined landscape of plausible failure scenarios \citep{Leveson2011SafetyCases}.\\

Risk models can be used as inputs in a safety case. For instance, in a safety case arguing that a system is sufficiently safe to deploy because ``all key hazards have been identified, estimated and mitigated'', risk modeling outputs (i.e., risk scenarios and risk estimations) can be part of the evidence, notably to back the statement that ``all key hazards have been identified and estimated''. Yet, a safety case can also rely on other forms of evidence, such as testing, formal verification, or adherence to safety standards \citep{carlan2024dynamic}. These do not directly map to a specific event sequence, which makes the elucidation of specific failure pathways less obvious and direct.\\

All in all, safety cases should not be used as a substitute for thorough risk modeling, but rather as another (potentially complementary) tool in the risk management toolbox.

\medskip
\footnotesize
\textit{Note:} Alongside the risk of argumentative closure masking parts of the risk landscape, \citet{Leveson2011SafetyCases}'s critique of safety cases also insists that they encourage confirmation bias, with safety cases’ authors focusing on evidence proving system safety and unconsciously disregarding others. Leveson concludes that ``assurance cases'' would produce better results should they ``focus not on showing that the system is safe but on attempting to show that it is unsafe''—which would make them more akin to the scenario-building phase of thorough risk modeling.
\end{numberedbox}

any in the field thus appear to conceive of safety cases and the scenario building part of risk modeling as equivalent. However, \textbf{safety cases are not equivalent to comprehensive scenario building} (see \textbf{Box~\ref{box:safety-case}}). The two are better conceived of as complementary: for instance, the results of a risk modeling exercise can be used as input in a safety case. 

Distinct from the safety case approach, other research focuses on tools for direct scenario development. Convergence Analysis’s research program on “scenario planning” \citep{Convergence_ScenarioResearch} is an example. \citet{chin2025dimensional} proposes a scenario building methodology for catastrophic AI risks like CBRN, cyber offense, or loss of control. The proposed methodology combines "dimensional characterization" to systematically analyze risks across seven key dimensions (such as intent, competency, linearity, or reach) with "risk pathway modeling" to map out the step-by-step causal progressions from an initial hazard to a resulting harm. 

\citet{wisakanto2025adapting} suggest adapting techniques from probabilistic risk assessment in the nuclear or aerospace industry to AI. When it comes to scenario building, the paper advocates for the use of\textbf{ “aspect-oriented hazard analysis”} to systematically identify hazards through considering AI system “aspects” (capabilities, domain knowledge, and affordances). Once this is done, authors propose using \textbf{“risk pathway modeling”} to trace how AI aspects lead to real-world harms through causal sequences\footnote{These risk pathways consist of six elements: source aspects, source aspect-adjacent hazards, intermediate steps, propagation operators, terminal aspect-adjacent hazards, and terminal aspects.}. Complementary approaches include \textbf{“forward chaining”}, which is reminiscent of event-tree analysis, and \textbf{“backward chaining”}, beginning with potential harms and working backwards to identify causes, which is reminiscent of fault-tree analysis. \citet{wisakanto2025adapting} also distinguish between “competence-based hazards”, arising “when highly effective capabilities lead to harmful consequences” and “incompetence-based hazards”, stemming from system limitations or failures. Finally, in building risk scenarios, authors consider how \textbf{“propagation operators”}, such as accumulation, correlation, adversarial exploitation, and sociotechnical diffusion effect might amplify and transform risk.

Finally, there are some examples of authors applying traditional scenario building approaches to particular AI risks. This is the case of \citet{barrett2017model}'s attempt to model the major pathways to an Artificial SuperIntelligence (ASI) catastrophe, who notably use fault trees to identify “combinations of events and conditions that could lead to AI catastrophe”. While the paper does not quantify risks, the authors see their model as a “foundation for rigorous quantitative evaluation and decision-making on the long-term risk of ASI catastrophe”.

\subsection{Existing Research and Practices in AI Risk Quantification}

Current attempts at AI risk quantification are also nascent and diverse. Many efforts are limited to measuring model capabilities on specific benchmarks \citep{Anthropic_RSP_Announcement_2023, OpenAI2023_OurApproachToAISafety}. As noted in the introduction, while useful, capability scores are proxies for hazard, not measures of real-world risk, since they often miss crucial contextual factors like threat actor behavior or deployment environment.

More sophisticated approaches are emerging. \citet{perrier2025statistical}, for instance, proposes a framework for partitioning AI-related events into a multi-stage pipeline, modeling dependencies using Markov chains and copulas, and using "lookalike distributions" from other domains to handle data scarcity. \citet{rodriguez2025framework} evaluate how AI helps cyber attackers by: first identifying representative attack scenarios, then using bottleneck analysis to find the most critical attack steps, and finally measuring how much AI reduces the cost of executing those attacks.

Other work has explored quantification within safety cases. Researchers have proposed methods for assigning probabilities to claims and aggregating them to produce an overall confidence estimate \citep{clymer2024safety, balesni2024towards, barrett2025assessing} but this approach faces significant challenges. For example, achieving high confidence in a top-level claim requires extremely high confidence in every sub-claim, and simplistic aggregation methods often rely on problematic assumptions of independence among arguments \citep{balesni2024towards,barrett2025assessing}\footnote{Note that risk quantification using any factorized model does not solve that issue: for a top level result (e.g. expected impact), it is hard to get high confidence in an exact value without getting very high confidence in each of the leaf nodes below. However, risk quantification with Bayesian networks helps updating evidence for leaf nodes later to reduce the top level uncertainty. It also helps in dealing with confidence properly, as Bayesian Networks specifically allow for propagation and attribution of the uncertainty. In addition, while safety cases aim to prove that something is safe, thereby demanding certainty, risk models benefit from certainty, but are not constructed to need it in order to be useful.}. \citet{clymer2024safety} propose a safety case framework to combine evidence on threats and the effectiveness of mitigations (in their example, API-based safeguards) to produce an overall quantitative estimate of risk. 

\citet{wisakanto2025adapting} propose a semi-quantitative risk estimation approach using coarse-grained bands rather than precise probabilities, which they argue is more appropriate for "novel or low-probability, high-impact events where historical data is scarce." In practice, they suggest producing a 10 category \textbf{risk level matrix}, combining information on \textbf{harm severity levels} (from HSL-1, marginal to HSL-6, globally catastrophic) with information on \textbf{likelihood levels} characterizing the probability of occurrence with defined odds bands, from LL-0 to LL-8. For likelihood estimation, they suggest applying the following formula: "P(harmful scenario) = P(capability exists) × P(capability misused | exists) × P(harm occurs | misused)". 

\citet{Murray2025AIRisk} translates AI benchmark scores into risk estimates through expert elicitation. The study focuses on estimating the probability that a specific step of an AI enabled cyber attack risk model is achievable. Using the IDEA protocol \citep{hemming2018practical} for structured expert elicitation, they show cybersecurity experts increasingly difficult tasks that an LLM can solve (drawn from a cybersecurity benchmark), then ask: "If attackers had access to an LLM with this capability level, what would be their probability of successfully achieving the risk model step?" This creates a direct mapping from benchmark performance to real-world risk probabilities. \citet{Righetti2025DualUseBioterrorism} uses historical studies, expert elicitation and forecasting to convert capability evaluations into probability of occurrence and estimates of damage related to the risk of bioterrorism. 

\subsection{Going Forward: Promising Avenues to Consider}
The above review shows that despite progress, efforts remain fragmented between scenario building and risk quantification. \citet{wisakanto2025adapting}'s paper goes in the right direction by integrating scenario building and risk estimation into a coherent framework but most existing approaches still consider these two components separately. 

Moving beyond isolated capability evaluations and safety cases, the field should prioritize \textbf{building an integrated modeling process that tightly couples scenario building and risk quantification.} Scenario building should be conducted in a way that facilitates risk quantification, and risk quantification should build on the causal logic of scenario building. This involves using structured scenario-building techniques like Fault Tree or Event-Tree Analyses to map out causal pathways to harm. These structured scenarios then provide the logical foundation for quantification using dependency-aware methods like Bayesian Networks, allowing for a more systemic and comprehensive risk picture. Such an integrated process will help develop standardized quantitative measures of AI risk to compare risk levels between different systems.

Separately, quantification efforts would currently benefit from the \textbf{more rigorous use of expert judgment.} To address concerns about the limitations of current quantification \citep{goemans2024safety}, the focus should be on improving its credibility. This involves adopting rigorous structured elicitation protocols and explicitly reporting confidence levels attached to estimates. 

Finally, in adapting risk modeling to advanced AI, some of the characteristics of AI should be accounted for. \textbf{Advanced AI risk modeling must account for high-severity, low-probability ``tail risk''}. In practice, when resources (e.g. time) are constrained, risk modeling might end-up under-estimating potential catastrophic outcomes \citep{Haimes2004_Modeling_RiskAnalysis, Hendrycks2024_TailEvents_BlackSwans}. In the case of advanced AI, particular attention should be paid to avoiding this pitfall.   

Considering the pace of technological advancement, advanced AI risk modeling must also be \textbf{dynamic and iterative}. Modeling cannot be a one-time affair, but a continuous process, with risk models being regularly updated to reflect new data, emerging capabilities, and changes in the threat environment \citep{carlan2024dynamic}. Here, the proposal by \citet{carlan2024dynamic} for a “Dynamic Safety Case Management System (DSCMS)” offers valuable insights. A similar system for risk modeling could incorporate quantitative metrics with predefined thresholds—analogous to Safety Performance Indicators (SPIs) or the Key Risk Indicators (KRIs) mentioned in \citet{campos2025frontier}—to enable a continuous process of monitoring and updating risk modeling as new data becomes available. 
While regular updates are helpful, it should be noted that as AI capabilities increase, risk modeling might become increasingly difficult and possibly less efficient in producing real-world risk estimates, if models engage in routine sandbagging, or if new capabilities are not well captured in saturated benchmarks. This possible future limitations does not diminish the present-day relevance of risk modeling however. 

Another potential limitation is linked to the fact that foreseeing all the possible sequences of events that can lead to harm, in the case of a complex technology like AI, is very difficult. The general-purpose nature of AI makes exhaustive scenario building difficult. This means that \textbf{prioritization is key.}  An effective framework must employ structured techniques—such as the vulnerability and bottleneck analyses discussed in~\cref{subsubsec221:scenariobuilding} (e.g. FMECA, FTA to identify Minimal Cut Sets, or bottleneck analysis) to focus limited resources on the scenarios that contribute most significantly to the overall risk. 

\section{How Does Risk Modeling Fit In Risk Management in Other Safety-Critical Industries?}
\label{sec4:FiveIndustries}

 Risk modeling is commonly used in safety-critical industries. However, the precise context of its use varies from industry to industry. As explained in introduction, this reflects in part variations in risk tolerance between industries, which materialize in different way to manage risks, and different ways to measure and verify safety - in other words, to conduct \textit{safety analysis}. In particular, industries each use their own mix of \textbf{deterministic safety analysis (DSA)} and \textbf{probabilistic safety analysis (PSA)}. 

\textbf{Deterministic safety analysis aims to demonstrate that the system under review meets safety requirements under challenging conditions} – e.g. the worst initial operating state, delayed operator response, or additional equipment failures\footnote{Assuming the worst scenarios amounts to adding “safety margins” to account for uncertainties and unknowns. If the system can survive this pessimistic test, the logic goes, it should cope with more likely, less severe conditions. For example, in a nuclear reactor, a "Loss of Coolant Accident" (LOCA) – a sudden breach of the reactor cooling system – is a scenario considered. A deterministic safety analysis in that case must prove that even with a LOCA, and assuming the single worst failure of a safety system concurrent with it, the reactor’s emergency core cooling can still prevent core damage.}. It analyzes whether the system can withstand these failures without unacceptable consequences. It focuses on a limited number of predefined, credible accident scenarios (called "design basis accidents" or DBAs). The system’s response to these initiating events (e.g., a pipe breaking, or loss of coolant, in a nuclear power plant) is analyzed against fixed success criteria (e.g., maximum fuel temperature, radiation dose limits), without explicitly quantifying event probabilities. Deterministic analysis either relies on empirical stress-tests or uses established engineering principles and physical laws to predict the system's response and the consequences of the event. DSA produces binary outcomes: the system is either safe, if criteria are met, or unsafe \citep{de2019deterministic}. 

By contrast, \textbf{probabilistic safety analysis aims to draw a picture of the overall risk landscape.} It seeks to list all potential credible accidents that could lead to undesirable outcomes \citep{de2019deterministic} in order to estimate their likelihood and potential impact. Instead of relying on theoretical and/or empirical knowledge of systems to select credible accidents, PSA typically uses scenario building techniques allowing to think through as many potential accident sequences as possible. It then estimates their occurrence likelihood and severity using available historical data, expert judgment, or test results if available. Thousands of scenarios may be evaluated in a complex PSA, from high-frequency/low-consequence ones to low-frequency/high-consequence ones. Contrary to DSA, which purposefully uses pessimistic assumptions, PSA strives for realism, using best-estimate models and data to reflect true risk, trying to neither over- nor underestimate expected risk. The end result is not a binary safe/unsafe determination, but a quantitative risk profile: metrics like the annual probability of a core meltdown, or a frequency-severity curve of consequences \citep{pate1996uncertainties, keller2005historical}. 

\textbf{Risk modeling is applied slightly differently in a deterministic or probabilistic context.} In a deterministic safety analysis, scenario building efforts are focused on a pre-selected, bounded set of design basis accidents representing the most severe failure modes deemed credible. Risk estimation usually does not imply quantifying likelihood and harm, but rather assessing whether the system still performs adequately when these failure conditions occur, using theoretical knowledge of the system or observed behavior. In a probabilistic safety analysis, scenario building involves exploring a wide range of potential failure pathways and their inter-dependencies. Risk estimation focuses on assigning probabilities and estimation of harm to initiating events and subsequent failures to derive an overall probabilistic measure of risk.

\textbf{In a regulatory context, the end product of risk modeling is also used in a slightly different manner depending on whether the approach is deterministic or probabilistic.} DBA scenarios and associated risk estimation results can be used directly as deterministic safety criteria in e.g. a facility certification context. By contrast, the probabilistic risk profile of a facility must be compared to a predefined risk tolerance; the probability of occurrence of a particular risk model could also be used as evidence in a safety case to determine whether the facility meets safety requirements. 

\subsection{Risk Modeling in the Nuclear Industry}

The nuclear industry pioneered the large-scale application of probabilistic risk modeling. The landmark 1975 Reactor Safety Study (WASH-1400) was the first to systematically apply a probabilistic approach, integrating fault trees and event trees to model complex accident pathways and quantify their likelihood and consequences \citep{NUREGKM0010_2016}. The impact of this probabilistic approach expanded significantly after the 1979 Three Mile Island accident, which involved a cascade of interacting equipment failures and human errors—a scenario that traditional deterministic analysis had struggled to predict, but which the new probabilistic methods were well-suited to model \citep{keller2005historical}. This event validated the need for a modeling approach that could capture complex system interactions and identify major risk contributors, enabling a more targeted allocation of safety resources.

However, \textbf{the industry’s risk management approach currently relies on a combination of both probabilistic and deterministic safety analyses.} 

First, a probabilistic approach is often used and comprehensive, quantitative risk modeling is conducted. Scenario building in \textbf{nuclear probabilistic risk analysis (PRA)} is very comprehensive. It uses event trees to map out potential accident sequences following an initiating event and fault trees to analyze the failure probabilities of the safety systems within those sequences. This allows for the modeling of thousands of scenarios. Risk estimation in nuclear PRA assigns frequencies and probabilities to initiating events and component failures—drawn from historical data, component testing, and expert judgment—to produce a quantitative risk profile. Contemporary PRA in the nuclear industry is structured on three levels: Level 1 estimates the frequency of reactor core damage; Level 2 assesses the probability of containment failure and radioactive release; and Level 3 models the off-site consequences to public health and the environment \citep{de2019deterministic}.

Second, \textbf{alongside PRA, national regulators, following standards established by e.g. the International Atomic Energy Agency (IAEA)} \citep{IAEA_SSR2_1_Rev1_2016} \textbf{also mandate DSA in the nuclear industry}, to ensure a baseline of resilience. Scenario building in nuclear DSA is intentionally bounded. It focuses on a pre-defined set of severe, credible Design Basis Accidents, such as a major coolant pipe break. As described above, risk estimation is not probabilistic. It assesses whether the plant's safety systems can meet pre-defined, conservative acceptance criteria (e.g., maximum fuel temperature) under worst-case assumptions. The outcome is a binary pass/fail judgment \citep{IAEA_SSR2_1_Rev1_2016}.

These two modeling methodologies are complementary and form a core part of the industry's \textbf{"defense-in-depth"}\footnote{Defense-in-depth is a core principle in safety engineering that involves implementing multiple, independent layers of protective mechanisms, such that even if one layer fails, others are still in place to prevent or mitigate the consequences of a failure. These layers are designed to be diverse, encompassing a range from physical barriers (like a nuclear reactor's containment building) and automated safety systems (like emergency cooling pumps) to administrative controls and human procedures (like operator training and emergency response plans). The goal is to create a highly resilient system that does not rely on any single component or safeguard being perfect.} philosophy. DSA provides a robust, non-negotiable safety baseline by proving the design is resilient against specific, severe challenges. PRA then offers a more comprehensive and realistic picture of the overall risk profile, identifying system interactions and vulnerabilities. \textbf{This dual-approach ensures risk is managed through both conservative design principles\footnote{Including defense-in-depth protocols and built-in safety margins applied throughout the analysis to account for uncertainties.} and a quantitative understanding of risk probabilities} \citep{NRC_DefenseInDepth_Glossary}.

\subsection{Risk Modeling in the Aviation Sector}

Similar to the nuclear industry, the aviation industry today employs a "defense-in-depth" approach, embedding multiple, layered protections to ensure system robustness against a wide range of hazards.
The global civil aviation industry's approach to safety underwent a significant transition from the late 1990s through the 2010s, moving from an approach built on deterministic requirements to one centered on Safety Management Systems (SMS), using a probabilistic approach. This also marked a shift from a rule-based approach to a more goal-based and process-oriented one. Instead of only demonstrating compliance with specific rules, manufacturers must now proactively identify hazards and manage their risks, using probabilistic methods to assess likelihood and severity and applying principles like \textbf{ALARP} (As Low As Reasonably Practicable)\footnote{ALARP is a principle in risk management stating that risks should be reduced to a level that is "as low as reasonably practicable." This means that mitigations should be implemented until the cost (in terms of money, time, or trouble) of further reduction becomes grossly disproportionate to the safety benefit gained.} \citep{Leveson2011SafetyCases,lee2006risk}. 

This international shift was driven by the International Civil Aviation Organization (ICAO) and by the recognition that simple compliance with deterministic design standards was insufficient to prevent "organizational accidents" involving complex interactions between human factors, procedures, and technology \citep{Wojcik1989PRA_AviationSafety,FAA_SMS_Initiative}. The publication of ICAO's first Safety Management Manual in 2006 (see updated version ICAO, \citeyear{ICAO_Doc9859_2018} )provided key guidance, and the shift was formalized globally when ICAO's Annex 19, "Safety Management," became applicable in 2013, making SMS a required international standard \citep{Skybrary_ICAO_Annex19}. 

\textbf{Today, risk modeling in international civil aviation blends probabilistic and deterministic techniques.} The ICAO provides guidance for using a semi-quantitative risk matrix (see~\cref{Figure1}) to estimate the severity and probability of various hazard scenarios in its Safety Management Manual (Doc 9859)  \citep{ICAO_Doc9859_2018}. Concurrently, stringent deterministic requirements remain central to aircraft design and certification. A classic case is the requirement to demonstrate that an aircraft can still safely climb after an engine failure at the most critical point during takeoff \citep{lee2006risk}. This single-failure criterion is deterministic: the survivability of this worst-case event must be guaranteed by the design, regardless of the failure's low probability.

\subsection{Risk Modeling for Cybersecurity}

Scenario building in the cybersecurity field often takes the form of \textbf{threat modeling} (e.g. OWASP \citeyear{OWASP_STRIDE_ThreatModeling}), a process used to scope and structure the scenario space (assets, trust boundaries, plausible attacker actions) by adopting an attacker's point of view. Frameworks such as STRIDE (OWASP \citeyear{OWASP_STRIDE_ThreatModeling}) classify threats against data-flow diagrams, while PASTA (Process for Attack Simulation and Threat Analysis \citep{UcedaVelez2015_RiskCentricThreatModeling_PASTA}) offers a staged, attacker-centric process that links technical threats to business impact\footnote{STRIDE classifies threats into six categories—Spoofing, Tampering, Repudiation, Information Disclosure, Denial of Service, and Elevation of Privilege—while PASTA is a seven-stage, attacker-centric method: define objectives; define technical scope; decompose the application; analyze threats; analyze vulnerabilities/weaknesses; model attacks; assess risk \& impact \citep{MSLearn_TMT_Threats_2022, VerSprite_PASTA_whitepaper}.}. 

While scenario building techniques are well-developed, data scarcity and rapid change are major issues complicating risk modeling in the cybersecurity field. A systematic review by \citet{eling2020cyber} highlighted a serious “lack of available data on cyber risk,” especially regarding the frequency and severity of rare but costly events, as many organizations do not publicly share breach data - and of course, neither do attackers regarding their attempts. Moreover, cyber risks are highly interdependent: as attacks can cascade across connected systems, a risk model considering a particular risk in a particular system should take into account the possibility of propagation to other systems. Another difficulty comes from the adversarial nature of cyber risks: attackers constantly adapt to new defenses, making models built on past data quickly obsolete.

As a result of these challenges, cybersecurity risk modeling often starts by analyzing known vulnerabilities. These analyses frequently employ deterministic-style, severity-focused tools. The widely used \textbf{Common Vulnerability Scoring System (CVSS)} \citep{FIRST_CVSS_v4_0_Page}, for instance, provides a semi-quantitative score based on a vulnerability's intrinsic characteristics, largely setting aside the specific likelihood of it being exploited in a given environment. A more advanced deterministic technique, originally from system safety engineering and increasingly applied to cybersecurity since the mid-2010s is \textbf{System-Theoretic Process Analysis for Security (STPA-Sec)}. It examines the entire system for unsafe interactions and emergent properties that could lead to security breaches, going beyond traditional component-level failure analysis to identify novel hazards \citep{young2014integrated,abdulkhaleq2015comprehensive}.

\textbf{However, there is a strong push, particularly from the cyber insurance industry, for more rigorous quantification that can express risk in financial terms} \citep{sheehan2021quantitative, mukhopadhyay2019cyber, cremer2022cyber}. The \textbf{Factor Analysis of Information Risk} (FAIR) \citep{FAIRInstitute_WhatIsFAIR} framework is a leading methodology in this space. FAIR structures cyber risk into quantifiable factors and uses Monte Carlo simulations to estimate risk in monetary terms (e.g., “\$X million expected loss per year”), allowing it to be managed like other business risks. While powerful, applied versions of FAIR often rely on statistical approximations to remain computationally tractable, which can introduce inaccuracies, for example assuming light-tailed or bounded distributions for elicited inputs, independence between factors—choices that can understate tail losses and cross-system or accumulation risk if dependencies and fat-tails are present \citep{Jones2023_CRQ_FAIR}.

To address these limitations, researchers propose enhancing such frameworks with \textbf{Bayesian Networks} (BNs). BNs offer greater flexibility by integrating probabilistic and causal analyses, allowing for a more granular model of attack scenarios and better integration of expert judgment about attacker motivations and capabilities \citep{wang2020bayesian}. Note that the combination of FAIR with BNs exemplifies the critical integration of scenario building and risk quantification discussed above: it uses structured threat scenarios as the direct foundation for its probabilistic evaluations, demonstrating that a coherent scenario is a prerequisite for robust quantification.

\subsection{Risk Modeling in the Financial Sector}
 
Risk management in the financial sector is heavily influenced by the concept of risk appetite—a quantitative statement of how much risk (e.g. probability of loss beyond a certain threshold) an institution is willing to accept. This drives a strong emphasis on probabilistic risk modeling to produce loss likelihood estimates and ensure that exposures remain within the predefined limits corresponding to the risk appetite.

Two central concepts underpin financial risk quantification: \textbf{Value-at-Risk} (VaR) and \textbf{Conditional Value-at-Risk} (CVaR). VaR estimates the maximum potential portfolio loss over a specific time frame at a given confidence level. For example, if a portfolio has a one-day 95\% VaR of \$1 million, it means there is a 95\% chance that the portfolio will lose no more than \$1 million in a single day. However, VaR says nothing about what happens in the worst 5\% of cases. CVaR, also known as Expected Shortfall, extends this by measuring the average loss beyond the VaR threshold. CVaR would be the average loss on those days when the portfolio's losses exceed the \$1 million VaR, providing a much better measure of extreme tail risks. 

Market risks are often analyzed with Monte Carlo simulations. Operational risks, arising from internal failures or fraudulent activities, are quantified using \textbf{loss distribution approaches (LDA)}, which involves statistically modeling the expected frequency and severity of potential operational losses.

\textbf{While probabilistic techniques dominate, deterministic approaches are also used, particularly for regulatory compliance.} Regulators—i.e., prudential supervisors such as the U.S. Federal Reserve or the EU’s EBA/ECB-SSM, following global overarching guidelines from e.g. the Basel Committee’s Stress testing principles \citep{BCBS_StressTesting_Principles_2018}—mandate periodic stress tests that apply fixed, severe-but-plausible scenarios, such as a sharp market downturn or a severe recession, to assess resilience. These deterministic analyses use fixed, worst-case assumptions to ensure a baseline of stability\footnote{Banks also use deterministic tests to ensure the safety of critical, non-negotiable assets whose failure is not an option, such as their core payment systems—the essential infrastructure for processing all transactions from ATM withdrawals to interbank transfers.}. 

\subsection{Risk Modeling in Submarine Operations}

Due to the catastrophic potential of failures, submarine operations—particularly the U.S. nuclear Navy’s \textbf{SUBSAFE} program—prioritize a highly deterministic safety culture. SUBSAFE’s overarching objective is to provide \textbf{maximum reasonable assurance (MRA)} that submarine hulls remain watertight and that the boat can recover from flooding, a certification stance anchored in \textbf{non-negotiable requirements} and demonstrable conformance rather than optimization trade-offs. Institutionally, SUBSAFE is embedded in Navy oversight/training and maintenance doctrine, making conformity with it a formal \textbf{certification} function rather than a discretionary engineering choice \citep{Leveson2012_SUBSAFE}. 

\textbf{Objective Quality Evidence (OQE)} is the program’s evidentiary foundation and is defined as any statement of fact, quantitative or qualitative, about product/service quality based on observations, measurements, or tests \textbf{that can be verified.} Because probabilistic risk assessments cannot be verified, they are \textbf{not used for certification} \citep{depetro2021fire,Leveson2012_SUBSAFE}. Practically, OQE must demonstrate that deliberate steps were taken to comply with requirements—and \textbf{without OQE there is no basis for certification}, regardless of who did the work or how well they did it. SUBSAFE further emphasizes institutional separation of powers—often described as a \textbf{“three-legged stool”}—to keep design, materials/parts control, and fabrication/testing (and their documentation/traceability) independently checkable as sources of OQE.

In terms of risk modeling, this deterministic philosophy shapes both scenario building and risk estimation. Scenario building begins with Hazard Identification (HAZID) to list credible hazards (e.g., onboard fires, critical system failures), then uses fault and event trees to model pathways to catastrophic failure. The \textbf{risk estimation step is not probabilistic:} instead of assigning likelihoods, engineers must produce \textbf{OQE that the system can withstand the scenario} (e.g., demonstrated hull strength at specified depth/pressure; verified material pedigree and testing). This \textbf{assurance-by-evidence} approach operationalizes the MRA goal in day-to-day certification and maintenance activities (e.g. material control, traceability, and recurring training/qualification across the workforce).

This does not mean probabilistic methods are absent. They are used selectively where no historical OQE exists—for example, assessing hazards from \textbf{new technologies}. For example, \citet{depetro2021fire} use a semi-quantitative approach with Bayesian techniques to assess fire scenarios linked to new lithium-ion battery systems, estimating their likelihood to help inform design modifications. Core certification still depends on accumulating OQE to meet SUBSAFE requirements but probabilistic analysis is used selectively to model novel risks characterized by high epistemic uncertainty.

\section{How could Risk Modeling Be Used in the Context of Advanced AI Risk Management?}
\label{sec5:AdvancedRiskManagement}

Two lessons can be derived from the above review of risk modeling in five safety-critical industries: 

The \textbf{first} is that the use of risk modeling is \textbf{often mandated by national regulators} as part of risk management requirements, and the way in which it is practiced in a particular industry is often \textbf{specified in international standards developed by international institutions.} Whether they are flying passengers, lending money, or building nuclear reactors, safety-critical industries routinely engage in risk modeling aligned with standards developed by the ICAO, the Basel Committee or the IAEA.

As explained in introduction, questions related to the overall shape of risk management, which ultimately partly hinge on risk tolerance preferences, are beyond the scope of the present paper.
The above review of literature demonstrates the influence of answers to these questions on the practice of risk modeling in particular industries. To develop a mature practice of advanced AI risk modeling, questions related to responsibility sharing between e.g. international organizations, national regulators, and industry will have to be tackled. These notably include:  
\textit{Is risk modeling mandatory? If so, who mandates it? 
Do particular guidelines/standards have to be followed? Who sets them? 
Should there be an equivalent to ICAO/ IAEA for AI? 
Who audits risk modeling? 
Who is responsible for conducting risk modeling? 
Should the results be public?  }

Other questions related to how AI risk modeling should fit into AI risk management are directly related to the choices made in this first overarching set of questions on the shape of risk management. For example, the outputs of risk modeling could be primarily used as inputs to safety cases or to be evaluated against predetermined risk tolerance thresholds by regulators delivering deployment licenses. This choice will reflect the favored sharing of responsibility between regulators and industry in risk management. 

\begin{figure}
  \centering
  \fbox{\includegraphics[width=12cm,height=8cm,keepaspectratio]{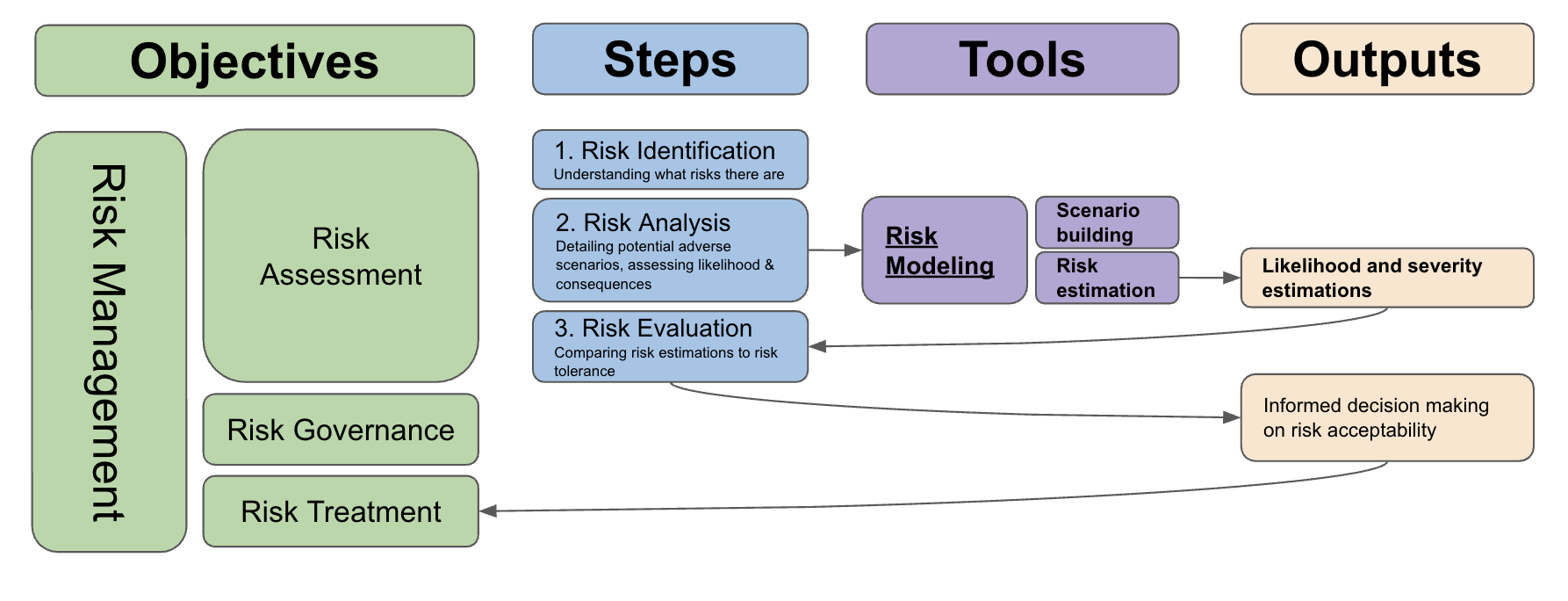}}
  \caption{A potential framework for integrating risk modeling into risk management~\citep{campos2025frontier}}
 \label{Figure2}
\end{figure}

As one potential example of how advanced AI risk modeling \textit{could} fit into a broader risk management context, \cref{Figure2} above presents the architecture proposed in \citet{campos2025frontier}. Drawing on established practices from other industries, the authors propose a framework\footnote{Key building blocks include risk identification, risk analysis \& evaluation (assessing the likelihood and severity of risks and comparing them to risk thresholds), risk treatment (mitigating the risks), and risk governance (implementing an institutional set-up which guarantees that risk management is done effectively, transparently and with appropriate checks and balances).} in which risk modeling is used by AI developers to produce real-world risk estimates of the harm associated with an AI system, expressed in terms of e.g. economic damage, or number of lives affected, combined with the probability that this harm materializes. Regulators are then charged with comparing these estimates to predetermined risk tolerance levels, to assess whether the AI system should be deployed or not.

The \textbf{second lesson from a look at these five industries} is that \textbf{risk management in safety-critical industries always mixes probabilistic and deterministic elements, and that mix influences their practices of risk modeling.} The balance of the blend varies. For example, in the nuclear and aviation sectors, safety rests on a deterministic baseline designed to ensure resilience against specific scenarios; probabilistic assessment adds a comprehensive view of the full risk landscape, including complex system interactions. In the financial sector, risk modeling is dominated by probabilistic quantification, with deterministic stress tests serving as a regulatory backstop. Conversely, submarine operations, with their emphasis on zero-failure tolerance, prioritize deterministic risk modeling, using probabilistic methods only selectively to assess novel technologies less amenable to deterministic testing.

Discussion of the particular blend of deterministic vs. probabilistic elements which would be adapted to advanced AI involves considerations of risk preferences and cannot be properly addressed here either. However, considering that in \textit{every industry} reviewed, at least some deterministic elements are used to \textit{ensure} that very likely and/or very consequential risk scenarios are avoided, \textbf{this paper makes the claim that a functional risk modeling approach for advanced AI should also mix probabilistic and deterministic elements}. We argue that some risks have a severity of associated harm so high that that the probability of their occurrence \textit{must} be zero. We contend that mandating a deterministic safety approach, with deterministic safety proofs, for these risks, would be aligned with common practices in other industries. 

However, in the case of advanced AI, ensuring at least some determinism in risk modeling and safety practices is highly complicated by the fact that AI systems do not comprise of components that can be independently evaluated. As mentioned above, in the current paradigm, AI development is not primarily guided by a knowledge of theoretical principles governing its function—akin to the physical laws of a nuclear reactor—but largely by the empirical observation that scaling compute and data yields more capable models. This entails profound epistemic uncertainty about a model's internal logic. This "grown" nature invalidates a core assumption of deterministic safety analysis, which relies on pre-selecting a finite list of "design basis accidents" and on proving that the system can withstand them. For advanced AI, this is unworkable for several reasons: 
\begin{itemize}
    \item the vastness of potential behaviors due to frontier models being general-purpose systems deployable across countless domains, the emergence of novel capabilities (the "unknown unknowns")~\citep{ganguli2022predictability}, and our limited understanding of AI models, which make it extremely \textbf{difficult to define a complete and stable set of credible accident scenarios}~\citep{IntAIsafety2025};
\end{itemize}
\begin{itemize}
    \item the fact that AI models can actively find ways around safety rules (through e.g. "specification gaming"), meaning that even a \textbf{well-defined safety boundary may not be robust};
\end{itemize}
\begin{itemize}
    \item the nature of some catastrophic tail risks, such as loss of control, which cannot be directly tested empirically. Identifying precisely which observables of AI systems are indicative of catastrophic risk or its frequency is very difficult. This means \textbf{AI safety is not currently verifiable with the certainty of traditional engineered systems required by a purely deterministic regime}.
\end{itemize}

 Following from the above, this paper claims that \textbf{to complement and enable a robust risk modeling practice in advanced AI, research in verifiable AI safety is critical}.\footnote{Incidentally, the point after which one would need to enforce deterministic guardrails, in terms of harm severity, is likely to coincide with the AI capability point, mentioned above, after which risk modeling might become less efficient, because models cannot be properly evaluated anymore (because benchmarks have been saturated, or because models are routinely sandbagging).} Investing in research for \textbf{provably safe AI architectures} could lead to producing the kind of deterministic guarantees that are currently unavailable. The realization that AI is \textit{not currently verifiable} is, in fact, at the heart of some research agendas geared at building guaranteed safe AI, which fundamentally aim to improve the deterministic verifiability of AI safety \citep{dalrymple2024towards, petrie2025flexible}. Interpretability research could offer a complementary path by potentially making current architectures' decision-making transparent enough to verify safety properties \citep{bereska2024mechanistic}. \textbf{For advanced AI risk management to be on par with standards upheld in other industries, these agendas should be urgently enhanced.}

\section{Conclusion}
\label{sec6:conclusion}
This paper argued that \textbf{sound advanced-AI risk management requires sound advanced-AI risk modeling}. Risk modeling is essential to understand and manage the complex risks arising from AI, yet it remains underused. Across safety-critical industries—from nuclear power to submarine operations—risk modeling is foundational to evidence-based decisions about complex technologies. For advanced AI, adopting an analogous practice is necessary to build a rigorous, transparent, and accountable risk-management regime that aligns deployment decisions with socially acceptable risk.

We defined \textbf{risk modeling as the joint exercise of scenario building and risk estimation}, jointly necessary for decision-making under epistemic uncertainty. At the technical level, we showed how risk modeling foundational concepts, scenario-building tools, and quantitative techniques can be adapted to advanced AI. We reviewed emerging AI-specific practices and highlighted gaps to be filled, complementing disconnected capability evaluations and safety cases. Promising avenues include building a \textbf{modeling process in which quantitative estimation builds on detailed causal scenarios}; ensuring this process is \textbf{iterative, to keep pace with technological change}; and \textbf{strengthening quantification through the principled use of expert judgment}.

To inform the role that AI risk modeling should play within risk management, we examined its use in five mature safety-critical industries. This allowed us to show that \textbf{defining the role of AI risk modeling in risk management implies choices related to responsibility sharing between industry and (international and national) regulators,  and ultimately to risk tolerance}. In one possible scenario, risk modeling could be mandated by national regulators, conducted by industry following international guidelines, and used to assess whether new models respect a predetermined risk tolerance threshold in deployment certification contexts.

Our survey of industry also helped showing how \textbf{risk modeling supports both deterministic and probabilistic analyses to manage risk at scale across sectors}. Based on this finding, \textbf{this paper argues that a functional risk modeling approach for advanced AI should also mix probabilistic and deterministic elements. We contend that mandating a deterministic safety approach for some of the highest severity AI risks would be aligned with common practices in other industries.} 

The paper’s original contributions are: a precise operationalization of “AI risk modeling” as the tight coupling of causal risk pathways with dependency-aware quantitative estimation; a blueprint to adapt risk modeling concepts, scenario building tools and quantitative estimation techniques to the case of AI; a review of emerging practices including a clarification of the role and limits of safety cases; a cross-sector synthesis of the role of risk modeling; a proposal for a governance-ready use of modeling; and a case for verifiable AI safety research as a risk management imperative given AI’s “grown, not designed” character.

Future work should prioritize three directions: 
\begin{itemize}
    \item First, further technical developments to sharpen methodology, including scalable, calibrated expert judgment; improved dependency and tail-risk methods; dynamic, iterative modeling with KRIs/KCIs; and validated mappings from lab capability evaluations to real-world risk.
\end{itemize}
\begin{itemize}
    \item Second, resolution of remaining subjective risk-management questions regarding responsibility sharing and risk tolerance.
\end{itemize}
\begin{itemize}
    \item Third, research into provably safe AI models to deliver the level of deterministic safety guarantees that is routine in domains built from first principles. Combined progress in these three strands of research would provide the stronger risk management apparatus that society expects for its most consequential technologies.
\end{itemize}

\section{Acknowledgment}

We would like to thank Adam Swanda, Aidan Homewood, Connor Stewart Hunter, Fabien Roger, and Luca Righetti for reviewing and providing valuable comments on this paper, as well as Max Schaffelder for his support with editing and visual content. All views expressed in this paper are our own, as are any potential remaining errors.

\newpage

\small
\bibliographystyle{abbrvnat}
\bibliography{references}
\pagebreak
\begin{appendix}
\section{Glossary}
\crefalias{section}{appendix}
\label{app:A}

\textbf{Bayesian Networks (BNs):} A type of graphical model that represents and quantifies probabilistic relationships among a set of variables. In a BN, nodes represent events or states, and connecting arcs represent conditional dependencies, making them well-suited for modeling complex causal chains and updating probabilities as new evidence becomes available. 

\textbf{Bottleneck Analysis:} A risk prioritization technique that focuses on identifying critical points or stages within a causal chain (such as a multi-step cyber attack) where an intervention would be most effective at disrupting the entire process, or where an AI-enabled capability would provide the greatest advantage.

\textbf{Defense-in-depth:} A core principle in safety engineering that involves implementing multiple, independent layers of protective mechanisms. The goal is to create a highly resilient system where the failure of a single layer does not lead to a catastrophic outcome, as other layers are still in place to prevent or mitigate the consequences.

\textbf{Deterministic Modeling / Deterministic Safety Analysis (DSA):} An approach to safety analysis that assesses a system's resilience against a pre-defined, bounded set of credible scenarios (called "Design Basis Accidents" or DBAs). Rather than calculating probabilities, it uses established engineering principles or stress-tests to determine if the system meets fixed success criteria, resulting in a binary (safe/unsafe) outcome.

\textbf{Event Tree Analysis (ETA):} A bottom-up, forward chaining scenario building technique that graphically maps the potential outcomes following a single initiating event. It explores the branching paths of possible consequences based on the success or failure of various safety functions or subsequent events. 

\textbf{Failure Mode, Effects, and Criticality Analysis (FMECA):} A forward chaining scenario building and risk prioritization technique that extends Failure Mode and Effect Analysis (FMEA). It involves identifying potential failure modes of components or processes, analyzing their effects on the system, and then ranking them by a criticality score, which is a function of their severity, probability of occurrence, and detectability. 

\textbf{Fault Tree Analysis (FTA):} A top-down, deductive scenario building technique where an undesired "top event" (a specific system failure) is traced backward to its root causes. It uses Boolean logic (AND/OR gates) to represent how combinations of lower-level failures can lead to the top-level outcome. 

\textbf{Harm:} The realized adverse outcomes resulting from a hazard. In the context of AI, this can include economic damage, loss of life, societal disruption, or other negative consequences. 

\textbf{Hazard:} The source of risk. In the context of AI, a hazard is often a model's capability, property, or tendency that has the potential to cause harm. 

\textbf{Minimal Cut Sets (MCS):} Derived from Fault Tree Analysis, an MCS is the smallest combination of component failures that, if they all occur, will cause the top-level system failure to occur. They represent the most direct pathways to failure and are critical targets for prioritization. 

\textbf{Probabilistic Modeling / Probabilistic Safety Analysis (PSA):} An approach to safety analysis that aims to identify and analyze as many potential credible accident scenarios as possible. It uses techniques like Fault Tree and Event Tree Analysis to model failure pathways and then assigns probabilities to each step to produce a quantitative risk profile (e.g., the annual probability of a specific failure), rather than a binary outcome. 

\textbf{Risk:} The combination of the probability of occurrence of harm and the severity of that harm. It is often conceptualized as a triplet: a scenario describing what can happen, the likelihood of that scenario, and its potential consequences. 

\textbf{Risk Scenario:} A logically laid-out sequence of causal steps linking a hazard (a source of risk) to a harm (a realized adverse outcome), taking into account the contexts in which the system may be deployed and the potential for intervening events or failures.

\textbf{Risk Tolerance:} A predefined level of risk that an organization, regulator, or society deems acceptable. In a risk management framework, estimated risks are compared against the risk tolerance to inform decisions about whether a system should be deployed or if further mitigation is required.
\end{appendix}

\end{document}